%% file: main.tex
\documentclass[letterpaper,11pt]{article} 

\input{our-preamble.tex}

\usepackage{verbatim}
\usepackage{titling}
\usepackage{cancel}
\usepackage{enumitem} 
\setlist{noitemsep,topsep=0pt,parsep=0pt} 
\usepackage[margin=1cm]{caption} 
\usepackage{subfig}
\usepackage{mathtools} 
\usepackage{multirow}
\usepackage[all]{xy}
\usepackage{tikz}
\usetikzlibrary{arrows,backgrounds,calc,fit,decorations.pathreplacing,decorations.markings,shapes.geometric}

\tikzset{every fit/.append style=text badly centered}

\numberwithin{equation}{section}
\usepackage[bookmarks=true,hypertexnames=false]{hyperref}
\hypersetup{colorlinks=true, citecolor=blue, linkcolor=red, urlcolor=blue}
\makeatletter

\newcommand{\Rmnum}[1]{\expandafter\@slowromancap\romannumeral #1@}
\makeatother

\newcommand{\Holant}{\operatorname{Holant}}

\newcommand{\Sym}{\operatorname{Sym}}

\newcommand{\trans}[4]{\ensuremath{\left[\begin{smallmatrix} #1 & #2 \\ #3 & #4 \end{smallmatrix}\right]}}

\newenvironment{remark}{\medskip{\bfseries \noindent Remark:}}{\par\medskip}{\par\medskip}

\tikzstyle{internal} = [draw, fill, shape=circle]
\tikzstyle{external} = [shape=circle]
\tikzstyle{square}   = [draw, fill, rectangle]
\tikzstyle{triangle} = [draw, fill, regular polygon, regular polygon sides=3, inner sep=3pt]
\tikzstyle{pentagon} = [draw, fill, regular polygon, regular polygon sides=5, inner sep=2pt, minimum size=14pt]

\newcommand{\posints}{\bbZ_{>0}}
\usepackage[normalem]{ulem}

\begin{document}

\title{\textbf{Perfect Matchings, Rank of Connection Tensors and Graph Homomorphisms}\thanks{A preliminary version of this paper appeared in the Proceedings of SODA 2019~\cite{Cai-Govorov-2019}.}}

\vspace{0.3in}

\author{Jin-Yi Cai\thanks{Department of Computer Sciences, University of Wisconsin-Madison. Supported by 
NSF CCF-1714275.} \\ {\tt jyc@cs.wisc.edu}
\and Artem Govorov\thanks{Department of Computer Sciences, University of Wisconsin-Madison. Supported by 
NSF CCF-1714275. 
 } \thanks{Artem Govorov is the author's preferred spelling 
of his name,
rather than the official spelling Artsiom Hovarau.
 } \\ {\tt hovarau@cs.wisc.edu}}

\date{}
\maketitle

\bibliographystyle{plainurl}

\begin{abstract}
We develop a theory of graph algebras over general fields.
This is modeled after the theory developed by
Freedman, Lov\'asz and Schrijver in~\cite{Freedman-Lovasz-Schrijver-2007}
for connection matrices,
in the study of graph homomorphism functions over  real edge weight
and positive vertex weight.
We introduce connection tensors for graph properties.
This notion naturally generalizes the concept of connection 
matrices. 
It is shown that
counting perfect matchings, and a host of other graph properties
naturally defined as Holant problems (edge models), cannot be
expressed by graph homomorphism functions with both
complex  vertex and edge weights
(or even from more general fields). 
Our necessary and sufficient condition in terms of connection tensors
is a simple  exponential rank bound. It shows that positive semidefiniteness
is not needed in the more general setting.
\end{abstract}

\thispagestyle{empty}
\clearpage
\setcounter{page}{1}

\input{introduction}

\input{preliminary-short}

\input{graph-algebras}

\input{applications-merged}

\input{main-theorem-proof}

\input{extensions}

\input{preliminary}

\input{appendix-previous-work}

\bibliography{references}

\end{document}

%% file: our-preamble.tex
\usepackage{amsmath, amsfonts, amssymb, amsthm, amscd, graphicx, 
enumerate, footmisc, mathtools, xcolor}
\providecommand{\U}[1]{\protect\rule{.1in}{.1in}}
\def\theenumi{\arabic{enumi}}

\def\theenumii{\alph{enumii}}
\def\p@enumii{\theenumi.}

\def\theenumiii{\arabic{enumiii}}
\def\p@enumiii{(\theenumi)(\theenumii)}

\def\p@enumiv{\p@enumiii.\theenumiii}
\newenvironment{lcases}
  {\left\lbrace\begin{aligned}}
  {\end{aligned}\right.}
\newcommand{\bbF}{\mathbb F}
\newcommand{\bbN}{\mathbb N}
\newcommand{\bbZ}{\mathbb Z}

\newcommand{\bbR}{\mathbb R}
\newcommand{\bbC}{\mathbb C}

\newcommand{\pmatch}{\operatorname{pm}}
\newcommand{\edgecol}{\operatorname{ec}}
\newcommand{\vdccover}{\operatorname{vdcc}}
\newcommand{\wmatch}{\operatorname{wm}}
\newcommand{\Char}{\operatorname{char}}

\newcommand{\rank}{\operatorname{rank}}
\newcommand{\symrank}{\operatorname{rk_\mathsf S}}
\newcommand{\matrank}{\operatorname{rank}}
\newcommand{\im}{\operatorname{im}}
\newcommand{\Span}{\operatorname{span}}
\newcommand{\id}{\operatorname{id}}

\newcommand{\calPLG}{\mathcal{PLG}}
\newcommand{\calG}{\mathcal G}
\newcommand{\calK}{\mathcal K}
\newcommand{\calP}{\mathcal P}
\newcommand{\bfx}{\mathbf x}
\newcommand{\bfy}{\mathbf y}

\newcommand{\calM}{\mathcal M}

\newcommand{\bfv}{\mathbf v}
\newcommand{\bff}{\mathbf f}

\newcommand{\calI}{\mathcal I}

\usepackage{fullpage}
\usepackage{mathrsfs}
\newtheorem{theorem}{Theorem}[section]
\newtheorem{claim}[theorem]{Claim}

\newtheorem{corollary}[theorem]{Corollary}

\newtheorem{lemma}[theorem]{Lemma}
\newtheorem{proposition}[theorem]{Proposition}
\numberwithin{equation}{section}

%% file: introduction.tex
\section{Introduction}

Many graph properties can be described in the general
framework called \emph{graph homomorphisms}. Suppose $G$ and $H$
are two graphs.  A mapping from
the vertex set $V(G)$  to the  vertex set $V(H)$
is a graph homomorphism if every edge $\{u, v\}$ of $G$
is mapped to an edge (or a loop) of $H$.
For example, if $H$ consists of two vertices $\{0, 1\}$
with one edge  between them and a loop at $0$,
then a vertex map $\phi: V(G) \rightarrow \{0, 1\}$
is a graph homomorphism iff $\phi^{-1}(1)$ is an independent
set of $G$. As another example, if $H = K_q$ is  a clique
on $q$ vertices (no loops),
then a vertex map $\phi: V(G) \rightarrow \{1, \ldots, q\}$
is a graph homomorphism iff $\phi$ is a proper vertex coloring of
$G$ using at most $q$ colors.

A more quantitative notion is the so-called partition
function associated with graph homomorphisms.
 The idea is that
we can consider a fixed $H$ with vertex weights and edge weights,
 and aggregate  all graph homomorphisms from $G$ to $H$,
 in a sum-of-product expression called the
partition function. This expression is invariant under graph isomorphisms,
thus expressing a graph property of $G$. This is  a weighted
counting version of the underlying concept~\cite{Lovasz-1967,GH-book, Borgs-et-al-counting-GH}.

More concretely, if each vertex $i \in V(H)$ has weight $\alpha_i$
and each edge $\{i, j\}$ of $H$ has weight $\beta_{i,j}$ (non-edge has weight $0$),
then the partition function $Z_H(\cdot)$ determined by $H$ is
\begin{equation}\label{eqn:def-graph-homomorphismshomomorphisms-pa-fn}
Z_H(G) = \hom(G, H) = \sum_{\phi: V(G) \rightarrow V(H)} 
\prod_{u \in V(G)} \alpha_{\phi(u)} 
\prod_{\{v, w\} \in E(G)} \beta_{\phi(v), \phi(w)}.
\end{equation}
The partition functions of graph homomorphisms
can express a broad class of weighted
counting problems.
Historically these  partition functions also arise in statistical
physics, where they play a fundamental role~\cite{baxter-6-8}. In classical physics
the  vertex and edge weights are typically (nonnegative) real numbers,
but in quantum theory they are complex numbers.
But even in classical physics, sometimes a generalization to
complex numbers allows a theoretically pleasing treatment.
E.g., Baxter generalized the parameters to complex values to 
 develop the ``commuting transfer matrix'' 
for the six-vertex model~\cite{baxter-6-8}. 
The book~\cite{GRS} (section 2.5.2)  treats the Hamiltonian of a 
one-dimensional 
spin chain as  an extension of the Hamiltonian of a 
six-vertex model with complex Boltzmann
weights.
 
Another source of fascination with these objects
comes from the classification program for counting problems
in complexity theory. In recent years many far-reaching classification
theorems have been proved classifying \emph{every} problem in a broad
class of counting problems as being either computable in polynomial time, or
being \#P-hard. This has been proved for 
graph homomorphisms (GH)~\cite{Dyer-Greenhill-2000, Dyer-Greenhill-corrig-2004,
Bulatov-Grohe-2005, Thurley-2009, Grohe-Thurley-2011, Goldberg-et-al-2010, Cai-Chen-Lu-2013},
for counting
constraint satisfaction problems (\#CSP)~\cite{bulatov-dalmu,
Bulatov, Dyer-Richerby, Cai-Chen-2017},
and for Holant problems~\cite{clx-holant,Backens-2017,Backens-2018,
glva-2013,Cai-Fu-Guo-W}. 
These
theorems are called complexity dichotomies. If we consider problem instances 
restricted to planar graphs and variables to take
Boolean values, there is usually a trichotomy,
where every problem is either (1) computable in polynomial time, or (2)
\#P-hard on general graphs but computable in polynomial time for 
planar graphs, or (3) \#P-hard on planar graphs.  Counting perfect
matchings, including weighted versions, is one such problem
that belongs to type (2). The planar tractability of
counting perfect
matchings is by Kasteleyn's algorithm (a.k.a.~FKT-algorithm)~\cite{Kasteleyn-1961, Kasteleyn-1967, Tempereley-Fisher-1961}.
Valiant introduced holographic algorithms to significantly extend
the reach of this methodology~\cite{Valiant-2004,Val06,cai-chen-book}. 
It is proved  that for
\emph{all}
 \#CSP where variables are Boolean
(but constraint functions can take complex values),
the methodology of 
holographic algorithms is \emph{universal}~\cite{Cai-Zhiguo-2017}. 
More precisely,
we can prove that
(A) the three-way classification above holds, 
and (B) the problems that belong to type (2) are precisely
those that can be captured by this single algorithmic approach,
namely a holographic reduction to Kasteleyn's algorithm.

\#CSP are
 ``vertex models'' where vertices are variables,
and constraints are placed on 
subsets of these variables. The partition function of GH
can be viewed as a special case of \#CSP  where each constraint is
a binary function as an edge weight
 (and for undirected graphs, a symmetric binary function).
When vertex weights are present they are unary functions.

In contrast to vertex models, one can consider ``edge models''
where each edge is a variable, and constraint functions are
placed at each vertex.  This is called a Holant problem~\footnote{Bal\'azs
Szegedy~\cite{B-Szegedy} 
 studied ``edge coloring models'' which are
equivalent to a special case of Holant problems where
for each arity $d$ a  symmetric vertex functions  $f_d$ 
is given and  placed at vertices of degree $d$.
In general, Holant problems allow different (possibly
non-symmetric) constraint functions from a  set  assigned at vertices; see~\cite{cai-chen-book}. 
}.
Counting perfect matchings is a Holant problem where the
constraint function at each vertex is the {\sc Exact-One} function.
Counting all matchings is a Holant problem with the
{\sc At-Most-One} constraint.
Other Holant problems include counting edge colorings, or vertex disjoint
cycle covers. Many problems in statistical physics, such as
(weighted) orientation problems, ice models,  six-vertex models etc.
are all naturally expressible as Holant problems.

It has been proved~\cite{Cai-Fu-Guo-W} 
that for Holant problems defined by an arbitrary
set of complex-valued
symmetric constraint functions on Boolean variables,
 (A) the three-way classification above holds,
but (B) holographic reductions to Kasteleyn's algorithm is
\emph{not} universal for type (2); there is an additional class of
planar P-time computable problems; these, together with 
holographic reductions to Kasteleyn's algorithm,
constitute a complete algorithmic repertoire for this class.
(It is open whether this also holds for 
non-symmetric constraint functions.)

But this should strike the readers as somewhat ironic.
Counting perfect matchings \emph{is} the problem that Kasteleyn's algorithm
solves for planar graphs. However this algorithmic
approach  is  proved universal for type (2)
only for vertex models but \emph{not} for edge models, and yet  
counting perfect matchings
is a \emph{quintessential} Holant problem.
It is most naturally expressed in the edge model.
It is \emph{not} naturally expressed  as a vertex model.

Or can it? 

Freedman, Lov\'asz and Schrijver~\cite{Freedman-Lovasz-Schrijver-2007} proved that
counting perfect matchings cannot be expressed as
the partition function of GH;
however their proof restricts to a definition of partition functions
with positive vertex weights and real valued edge weights.
More importantly they give a characterization for 
a graph property
to be  \emph{expressible} as such a partition function
of GH.

Their characterization consists of two conditions on
a \emph{connection matrix}:
 a rank condition
and a positive semidefiniteness condition.
But when we move from ${\bbR}$ to the complex field ${\bbC}$, this 
positive semidefiniteness condition breaks down.
At a high level, a succinct reason is  that  for complex matrices
$M$, it is \emph{not} true that $M^{\tt T} M$ is positive semidefinite. 
However, partition functions of GH
with complex weights are interesting~\cite{Lee-Yang-1952, Asano-1970, Newman-1974, Lieb-1981, GRS, baxter-6-8}, and natural
in the quantum setting.
More intrinsically (but less obviously), even if one is dealing with counting problems 
defined by real weights, 
complex matrices are essential as holographic transformations.
For example, the matrix
 $Z = \frac{1}{\sqrt{2}} \left[\begin{smallmatrix} 1 & 1\\
\mathfrak{i} & -\mathfrak{i} \end{smallmatrix}\right]$
is  one of the most important 
holographic transformations~\cite{cai-chen-book,
Cai-Fu-Guo-W} in dealing with orientation problems 
such as the six-vertex model, even when all given weights are real.
Note that $Z$ transforms  the binary {\sc Equality}
function to {\sc Disequality},
which is expressible in the form of signature matrices
as: $Z^{\tt T} I Z = \left[\begin{smallmatrix} 0 & 1\\
1 & 0 \end{smallmatrix}\right]$.
Thus the results in~\cite{Freedman-Lovasz-Schrijver-2007} do not answer whether counting perfect matchings,
and other similar problems
naturally expressible as Holant problems (edge models), can be expressed
as partition functions of GH
when
complex vertex and edge weights are allowed.
We note that Schrijver~\cite{Schrijver-2009,Schrijver-2013} gave
beautiful characterizations of  graph properties  expressible 
 as partition functions of GH
with complex edge weights but no vertex weights. Thus effectively
the vertex weights are all 1  (and there are also subtle differences in the 
model which we will discuss in subsection~\ref{appendix-previous-work}
in the  Appendix).
So the expressibility of these Holant problems as 
(\ref{eqn:def-graph-homomorphismshomomorphisms-pa-fn})
with complex $\alpha_i$ and $\beta_{i,j}$ remained an open problem.

In this paper we resolve this question.
We define the notion of a \emph{connection tensor}.
Then we give a 
tensor theoretic characterization of
when a graph parameter {can} be expressed as GH over \emph{any} field.
We show that there is \emph{only one} condition, which is both necessary and
sufficient for a graph parameter to be expressible by GH
with arbitrary vertex and edge weights,
and that is a simple exponential bound on the rank of the connection tensor.
Positive semidefiniteness is not required (and would not be
meaningful in a general field). 

This characterization is purely algebraic. 
As a consequence we show that
counting perfect matchings is not expressible as partition
functions of GH over an arbitrary field  (over a field of characteristic $p > 0$ we count perfect matchings modulo $p$).
We also prove the same inexpressibility for several other
naturally defined  Holant problems. Over  bounded degree graphs,
we prove a sharp threshold  for the domain size  ($|V(H)|$)
for expressibility, using holographic transformations.
While we dispense with
their positive semidefiniteness condition, the paper~\cite{Freedman-Lovasz-Schrijver-2007}
 is an inspiration for this work from which we borrow many
definitions and ideas.

To handle general vertex weights
a significant technical difficulty we have to overcome following
the approach of~\cite{Freedman-Lovasz-Schrijver-2007} is
 the possibility that vertex weights can cancel. In particular,
the possibility that the sum of all vertex weights can be $0$
creates a nontrivial
technical obstacle, and we have to introduce some consequential changes
to their proof.
To do that, in addition to the algebras of quantum graphs $\calG(S)$, 
we define a second type of algebras of quantum graphs $\calG_\subseteq(S)$,
where $S \subseteq \bbZ_+$ is a finite set of labels. 
In  $\calG(S)$ 
the generators are precisely $S$-labeled graphs, whereas in
 $\calG_\subseteq(S)$  
their label sets can be arbitrary subsets of $S$. We need to do this because 
the normalization argument from \cite{Freedman-Lovasz-Schrijver-2007} fails
in our setting
(precisely because the sum of all vertex weights can be $0$).
One technical  step 
involves correctly defining the notion of a projection 
from one quotient algebra to another,
$\hat \pi_S \colon \hat \calG \to \hat \calG_\subseteq(S)$. 
It must be onto $\hat \calG_\subseteq(S)$ which is not in general the 
same as $\hat \calG(S)$, the corresponding quotient algebra
without the normalization. (In~\cite{Freedman-Lovasz-Schrijver-2007} 
this  $\calG_\subseteq(S)$ was not needed.)
After the appropriate algebraic structures are all in place,
now somewhat more elaborate than that of~\cite{Freedman-Lovasz-Schrijver-2007},
we are able to establish our algebraic
characterization of expressibility of a graph property as GH.

An outline  of this paper
is as follows.
  Our main theorems are
Theorem~\ref{thm:uniform-exponential-rank-boundedness-converse} 
and 
Theorem~\ref{thm:uniform-exponential-rank-boundedness-converse-strengthened}.
To prove these theorems
we need a proper algebraic setting, and these are certain
infinite-dimensional \emph{algebras}, which are
vector spaces endowed with a multiplication.
These algebras are infinite-dimensional because
we wish to account for all finite labeled graphs in one structure.
Being infinite-dimensional introduces some technical complications.
In  Section~\ref{sec:Preliminary-short} we include
some basic notions, mainly regarding tensor spaces.
In the context of this paper the coordinates of these infinite-dimensional 
vector spaces represent partially labeled graphs 
in the algebra $\calG(S)$ or $\calG_\subseteq(S)$ (to be defined in Section~\ref{sec:proof-of-main-results}). 
In Section~\ref{section:graph-algebras}, 
we introduce the basic definitions of graph algebras 
and connection tensors of a graph parameter. 
In Section~\ref{sec:apps}, we show how the tensor theoretic
characterization can be used to prove that some graph properties cannot be 
expressed as GH over any field.
The main proof starts in Section~\ref{sec:proof-of-main-results}.
In subsection~\ref{subsec:monoid-and-algebra-of-graphs},
we define the monoid of partially labeled graphs and 
the algebra of quantum graphs in more detail.
We define the algebras of quantum graphs $\calG(S)$
and $\calG_\subseteq(S)$,
the ideals $\calK_S$ and $\calK_{\subseteq S}$,
and the respective quotients $\hat \calG(S)$ and $\hat \calG_\subseteq(S)$. 
We also introduce and prove the correctness of the definition 
of the aforementioned projection $\hat \pi_S \colon \hat \calG \to \hat \calG_\subseteq(S)$, 
which arises from the linear map $\pi_S \colon \calG \to \calG_\subseteq (S)$. 
As said before, the possibility that 
the vertex weights sum to $0$ does not allow us 
to perform the corresponding normalization step, and we cannot just simply 
repeat the proofs
from~\cite{Freedman-Lovasz-Schrijver-2007} without extending all the definitions systematically. 
With all the groundwork set, we may finally proceed to the main proof.   
We show the existence of the basis of idempotents in the quotient algebras $\hat \calG(S)$ 
for finite $S \subseteq \bbZ_+$ by constructing an isomorphism 
onto $\bbF^{r}$ for some $r$
(a composition of two isomorphisms,
see Lemma~\ref{lem:Phi(k)-tilde-calG[k]-isomorphism} 
and Corollary~\ref{cor:Phi(k)-hat-calG[k]-isomorphism}) 
thereby bounding their dimensions as well. After that we are able to proceed similarly to Section 4 
from \cite{Freedman-Lovasz-Schrijver-2007}, 
modifying the original proofs as needed.

Table~\ref{table:freq-used-notation} lists the main concepts and sets used in the paper.

\begin{table}[ht]
\centering
\begingroup
\setlength{\tabcolsep}{10pt} 
\renewcommand{\arraystretch}{1.5} 
\begin{tabular}{|c|}
\hline
$\bigoplus_\calI \bbF, \bbF^\calI, \Sym^n(\bbF^\calI), \symrank, T(f, k, n)$ \\
\hline
$\calPLG, \calPLG_\subseteq(S), \calPLG(S), \calPLG[k](= \calPLG([k]))$ \\
\hline
$\calG, \calG_\subseteq(S), \calG(S), \calG[k](= \calG([k]))$ \\
\hline
$\calK, \calK_{\subseteq S}, \calK_S, \calK_{[k]}$ \\
\hline
$\hat \calG, \hat \calG_\subseteq(S), \hat \calG(S), \hat \calG[k]( = \hat \calG([k]))$ \\
\hline
$\tilde \calG_\subseteq(S), \tilde \calG(S), \tilde \calG[k](= \tilde \calG([k]))$ \\
\hline
$U_S, U_\emptyset, U_k( = U_{[k]}), u_S = U_S + \calK, u_\emptyset = U_\emptyset + \calK$ \\
\hline
$\pi_S \colon \calG \to \calG_\subseteq(S), \hat \pi_S \colon \hat \calG \to \hat \calG_\subseteq(S)$ \\
\hline
\end{tabular}
\endgroup
\caption{\label{table:freq-used-notation}Main concepts and sets used in the paper.}
\end{table}

%% file: preliminary-short.tex
\section{Some Basic Concepts}\label{sec:Preliminary-short}

In this paper $\bbF$ denotes an arbitrary field, also
viewed as a one-dimensional vector space over $\bbF$. 
The  set $\bbF^\calI$ consists of tuples indexed by  $\calI$.
We let $\posints$ denote the set of positive integers.
For any integer $k \ge 0$, let $[k] = \{ 1, \ldots, k \}$. 
In particular, $[0] = \emptyset$.
For finite $\calI = [n]$ we write $\bbF^n$. By operations on components
$\bbF^\calI$ is an algebra (vector space and a ring).
By convention $\bbF^0 = \bbF^{\emptyset} = \{\emptyset\}$,
and $0^0 = 1$ in $\bbZ$, $\bbF$, etc.
We use $\bigsqcup$ to denote disjoint union.
In subsection~\ref{Multilinear-algebra-short}
we briefly state some concepts and results.
A more detailed account is given in Section~\ref{sec:Preliminary-continued} as an appendix.

\subsection{Multilinear algebra}\label{Multilinear-algebra-short}

We assume that the reader is familiar with tensors.
A main feature in this paper is that we deal with
infinite dimensional spaces and their duals;  and this infinite dimensionality
causes some  technical complications.
E.g., 
multilinear functions on $\prod_{i = 1}^n V_i$ can be naturally identified 
with the dual space $(\bigotimes_{i = 1}^n V_i)^*$ of 
linear functions on $\bigotimes_{i = 1}^n V_i$.
Moreover, $\bigotimes_{i = 1}^n V_i^*$ canonically embeds into 
$(\bigotimes_{i = 1}^n V_i)^*$ via  
$(\otimes_{i = 1}^n f_i) (\otimes_{i = 1}^n v_i)
=
\prod_{i = 1}^n f_i(v_i)$. 
A special case is that $(V^*)^{\otimes n}$ embeds into $(V^{\otimes n})^*$.
If all $V_i$'s are finite 
dimensional then this embedding is an isomorphism. 
However, if $V_i$
are infinite dimensional, this embedding is \emph{not} surjective.
To see this, consider $V^{\otimes 2}$ where $V$ is the linear span of
$\{e_i \mid i \in {\posints}\}$, where $e_i$ is 
the $0$-$1$ vector indexed by ${\posints}$ with a single $1$ at the $i$th position.
Let $f \in (V^{\otimes 2})^*$ be such that
$f(e_i \otimes e_j) = \delta_{ij}$, which is $1$ if $i=j$ and $0$ otherwise.
Then 
there is no tensor  $T \in  (V^*)^{\otimes 2}$
that embeds as $f$.
Indeed, any $T \in  (V^*)^{\otimes 2}$ is, by definition, a \emph{finite} sum
$T = \sum_{1 \le k <n} c_k f_k \otimes g_k$. If $T$ were to embed as $f$,
then consider the $n\times n$ matrix where
the
$(i,j)$ entry is $f(e_i \otimes e_j)$, which is the identity 
matrix $I_n$ of rank $n$. 
However $T(e_i \otimes e_j) =
  \sum_{1 \le k <n} c_k f_k(e_i) \cdot g_k(e_j)$, and
so the matrix for the embedded $T$ has rank $< n$,
being a sum of $n-1$ matrices of rank $\le 1$.

%
For any symmetric tensor
  $A \in \Sym^n(V)$ we define the symmetric rank of  $A$
to be the  least $r \ge 0$ for which $A$ can be expressed as
\[
A = \sum_{i = 1}^r \lambda_i \bfv_i^{\otimes n},
\quad \lambda_i \in \bbF, \bfv_i \in V,
\]
and we denote it by $\symrank(A)$.
If there is no such decomposition we define $\symrank(A) = \infty$. 
If $\symrank(A) < \infty$ then in any such expression of $A$
as a sum of $\symrank(A)$ terms
all $\lambda_i \ne 0$, all $\bfv_i \ne 0$ and are
pairwise linearly independent.
In subsection~\ref{sec:fin-sym-ten-rank} we show that if $\bbF$ is infinite,
then $\symrank(A) < \infty$ for all $A \in \Sym^n(V)$.
%
%
%
%
We prove all needed
 technical multilinear algebra statements in the Appendix (Section~\ref{sec:Preliminary-continued}).

\subsection{Weighted graph homomorphisms}

We recap the notion of weighted graph homomorphisms \cite{Freedman-Lovasz-Schrijver-2007}, but state it for an arbitrary field $\bbF$.

An ($\bbF$-)\textit{weighted graph} $H$ is a graph with a weight $\alpha_H(i) \in \bbF \setminus \{0\}$ associated with each node $i$ and a weight $\beta_H(i, j) \in \bbF$ associated with each edge $ij$. For undirected GH,
we assume $\beta_H(i, j)= \beta_H(j, i)$.

Let $G$ be an unweighted graph (with possible multiple edges, but no loops) and $H$ a weighted graph (with possible loops, but no multiple edges). 
A map $\phi \colon V(G) \to V(H)$ is a homomorphism if
every edge of $G$
goes to an edge or loop of $H$.
In this paper, it is convenient to assume that 
$H$ is a complete graph with a loop at all nodes by adding
all missing edges and loops with weight $0$. Then the weighted graph $H$ is described by an integer $q = |V(H)| \ge 0$ ($H$ can be the empty graph), a nowhere zero vector $a = (\alpha_1, \ldots, \alpha_q) \in \bbF^q$ and a symmetric matrix $B = (\beta_{i j}) \in \bbF^{q \times q}$.
In this setting every map  $\phi \colon V(G) \to V(H)$ is a homomorphism.
 We assign the weights
\begin{equation}\label{eqn:hom-phi-def}
\alpha_\phi = \prod_{u \in V(G)} \alpha_H(\phi(u)),
\quad \quad \quad
\hom_\phi(G, H) = \prod_{u v \in E(G)} \beta_H(\phi(u), \phi(v)),
\end{equation}
and define
\begin{equation}\label{eqn:hom-def}
\hom(G, H) = \sum_{\phi \colon V(G) \to V(H)}
\alpha_\phi \hom_\phi(G, H).
\end{equation}
When $G$ is the empty graph, i.e., $V(G) = \emptyset$, the only
map $\phi \colon \emptyset \to V(H)$ is the empty map $\phi = \emptyset$;
in that case we have the empty products
$\alpha_\emptyset = 1$, $\hom_\emptyset = 1$, and $\hom(G, H) = 1$.

If all node-weights and edge-weights in $H$ are $1$, then this is the number of homomorphisms from $G$ into $H$.
Without loss of generality we require 
all vertex weights $\alpha_H(i) \ne 0$ since any vertex $i$ with
 $\alpha_H(i) = 0$ 
can be deleted together with all incident edges $ij$ and loops at $i$.
%
%


Note that when $H$ is the empty graph, then
$\hom(G, H) = 0$ if $G$ is not the empty graph 
(because there is no map $\phi \colon V(G) \to V(H)$ in this case),
and $\hom(G, H) = 1$ if  $G$ is the empty graph
(because there is  precisely one 
 empty map $\phi = \emptyset$ in this case.)
The function $f_H = \hom(\cdot, H)$ is
a graph parameter, a concept to be formally defined  shortly.


%% file: graph-algebras.tex
\section{Graph algebras\label{section:graph-algebras}}


\subsection{Basic definitions}

An $\bbF$-valued \textit{graph parameter} is a function from
finite graph isomorphism classes to $\bbF$~\footnote{The concept can be defined over commutative rings but our treatment uses properties of a field.}.
 For convenience, we think of a graph parameter as a function defined on finite graphs and invariant under graph isomorphism. We 
allow multiple edges in our graphs, but no loops, as input to a graph parameter.
 A graph parameter $f$ is called \textit{multiplicative}, if for any disjoint union $G_1 \sqcup G_2$ of graphs $G_1$ and $G_2$ we have $f(G_1 \sqcup G_2) = f(G_1) f(G_2)$.


A $k$-labeled graph ($k \ge 0$) is a finite graph in which $k$ nodes are labeled by $1, 2, \ldots, k$ (the graph can have any number of unlabeled nodes). Two $k$-labeled graphs are isomorphic if there is a label-preserving isomorphism between them. 
We identify a ($k$-labeled) graph with its ($k$-labeled) graph isomorphism class.
We denote by $K_k$ the $k$-labeled complete graph on $k$ nodes, and by $U_k$, the $k$-labeled graph on $k$ nodes with no edges. In particular, $K_0 = U_0$ is the empty graph with no nodes and no edges.
A graph parameter on a labeled graph ignores its labels.

It is easy to see that for a multiplicative graph parameter $f$, either $f$ is identically $0$ or $f(K_0) = 1$. Every weighted graph homomorphism $f_H = \hom(\cdot, H)$ is a multiplicative graph parameter. 


The  \textit{product} of  two $k$-labeled graphs $G_1$ and $G_2$
 is defined as follows: we take their disjoint union, and then identify nodes with the same label. Hence for two $0$-labeled graphs, $G_1 G_2 = G_1 \sqcup G_2$ (disjoint union). Clearly, the graph product is associative and commutative with the identity $U_k$, so the set of all isomorphism classes of finite $k$-labeled graphs together with the product operation forms a commutative monoid which we denote by $\calPLG[k]$.


Let $\calG[k]$ denote the monoid algebra $\bbF \calPLG[k]$ consisting of all
finite formal linear combinations in $\calPLG[k]$ with coefficients from $\bbF$;
they are called ($k$-labeled, $\bbF$-)\textit{quantum graphs}.
This is a commutative algebra with $U_k$ being the multiplicative identity,
and the empty sum as the additive identity.
Later, in Section~\ref{sec:proof-of-main-results} we 
will expand these definitions to allow label sets
to be arbitrary finite subsets of $\posints$.

\subsection{Connection tensors}

Now we come to the central concept for our treatment. Let $f$ be any graph parameter. For all integers $k, n \ge 0$, we define the following $n$-dimensional array $T(f, k, n) \in \bbF^{(\calPLG[k])^n}$,
which can be identified with $(V^{\otimes n})^*$,
where $V$ is the infinite dimensional vector space with
coordinates indexed by ${\calPLG[k]}$, i.e., $V = \bigoplus_{\calPLG[k]} \bbF$.
The entry of $T(f, k, n)$ at coordinate $(G_1, \ldots, G_n)$ is $f(G_1 \cdots G_n)$; when $n = 0$, 
we define $T(f, k, n)$ to be the scalar $f(U_k)$. Furthermore, by the commutativity of the product the arrays $T(f, k, n)$ are symmetric with respect to its coordinates,
i.e., $T(f, k, n) \in \Sym(\bbF^{(\calPLG[k])^n})$.
Fix $f, k$ and $n$, we call the $n$-dimensional
 array $T(f, k, n)$ the ($k$th, $n$-dimensional)
 \textit{connection tensor} of the graph parameter $f$. When $n = 2$, 
a connection tensor 
is exactly
a \textit{connection matrix} of the graph parameter $f$ 
studied in \cite{Freedman-Lovasz-Schrijver-2007}, i.e., $T(f, k, 2) = M(f, k)$.

In contrast to~\cite{Freedman-Lovasz-Schrijver-2007}, we will be concerned with only one property of connection tensors, namely their symmetric rank.
The symmetric rank $\symrank(f, k, n) = \symrank(T(f, k, n))$, as a function of $k, n$, will be called the \textit{symmetric rank connectivity function} of the parameter $f$. This may be infinite, but for many interesting parameters it is finite, and its growth rate will be important for us.


\begin{remark}
In~\cite{Freedman-Lovasz-Schrijver-2007}, the matrix rank  of  $M(f, k)$
was used for the connection matrices. 
The results of the present paper use the symmetric tensor rank of the connection tensors $T(f, k, n)$ and hold for arbitrary fields. 
For $n = 2$, tensor rank coincides with
matrix rank, i.e.,  $\rank(T(f, k, 2)) =  \rank(M(f, k))$,
and furthermore if $\Char \bbF \ne 2$,
then the symmetric tensor rank also coincides, 
$\symrank(T(f, k, 2)) = \rank(T(f, k, 2)) = \rank(M(f, k))$.
Since the field in~\cite{Freedman-Lovasz-Schrijver-2007}
is  $\bbR$, the notions are consistent.
\end{remark}

\begin{proposition}\label{prop:graph-paramater-multiplicativity-criterion}
Let $f$ be a graph parameter that is not identically $0$. 
The following are equivalent:
\begin{enumerate}
\item\label{item-multiplicative-1}
$f$ is multiplicative.
\item\label{item-all-ge0-2}
$f(K_0) = 1$ and for all $n \ge 0$, $\symrank T(f, 0, n) = 1$.
\item\label{item-some-ge2-3}
$f(K_0) = 1$ and 
there exists some $n \ge 2$,  $\symrank T(f, 0, n) = 1$.
\end{enumerate}
%
\end{proposition}
\begin{proof}
Suppose $f \not = 0$ is multiplicative. 
Then $f(K_0)^2 = f(K_0)$, showing that $f(K_0) \in \{ 0, 1 \}$. If $f(K_0) = 0$, then the relation $f(G) = f(G) f(K_0)$ implies that $f(G) = 0$ for every $G$, which is excluded. So $f(K_0) = 1$.
Trivially $\symrank T(f, 0, n) = 1$ for $n = 0, 1$.
Fix any $n \ge 2$.
Then $f(G_1 \cdots G_n) = f(G_1) \cdots f(G_n)$ for any $0$-labeled graphs $G_1, \ldots, G_n$, which implies that $\symrank T(f, 0, n) = 1$.

Now suppose $f(K_0) = 1$ and  for some $n \ge 2$,  $\symrank(f, 0, n) = 1$.
This implies that there is a graph parameter $\phi$ and a constant $c_n$
 such that $f(G_1 \cdots G_n) = c_n \phi(G_1) \cdots \phi(G_n)$. 
Putting all $G_i = K_0$, we get $c_n \phi(K_0)^n = f(K_0) = 1$ 
so $\phi(K_0) \not =0$
and $c_n = 1 / \phi(K_0)^n$. 
Dividing $\phi$ by $\phi(K_0)$ we can assume that $\phi$ is normalized
 so that $f(G_1 \cdots G_n) = \phi(G_1) \cdots \phi(G_n)$ and $\phi(K_0) = 1$. Next, taking $G_1 = G$ and $G_i = K_0$ for $2 \le i \le n$ we see that $f(G) = \phi(G)$ for every $G$ and therefore $f(G_1 \cdots G_n) = f(G_1) \cdots f(G_n)$. Finally, substituting $G_i = K_0$ for $2 < i \le n$,  
we get $f(G_1 G_2) = f(G_1) f(G_2)$ so $f$ is multiplicative.
\end{proof}

\subsection{Connection tensors of homomorphisms}

Fix a weighted graph $H = (\alpha, B)$. 
Recall that 
in the definition of $\hom(\cdot, H)$ 
we assume
 $H$ to be a complete graph with possible
$0$ weighted edges and loops, but no $0$ weighted vertices.
For any $k$-labeled graph $G$ and mapping $\phi \colon [k] \to V(H)$, let
\begin{equation}\label{eqn:partial-hom-dec}
\hom_\phi(G, H) = \sum_{\text{\small $\substack{\psi \colon V(G) \to V(H) \\ \psi \text{ extends } \phi}$}} \frac{\alpha_\psi}{\alpha_\phi} \hom_\psi(G, H), 
\end{equation}
where $\alpha_\phi = \prod_{i \in [k]} \alpha_H(\phi(i))$,
and $\alpha_\psi$ and $\hom_\psi$ are defined by (\ref{eqn:hom-phi-def}).
Here $\psi  \text{ extends } \phi$ means that
 if $u_i \in V(G)$ is labeled by $i \in [k]$
then $\psi(u_i) = \phi(i)$,
so $\frac{\alpha_\psi}{\alpha_\phi}$
is the product of vertex weights of $\alpha_\psi$ \emph{not} in $\alpha_\phi$.
Then 
\begin{equation}\label{eqn:full-hom-dec}
\hom(G, H) = \sum_{\phi \colon [k] \to V(H)} \alpha_\phi \hom_\phi(G, H).
\end{equation}

Our main contribution in this paper is that a simple exponential bound in $k$ on the symmetric rank of the connection tensor of a graph parameter characterizes it being expressible as $\hom(\cdot, H)$.  This holds over all fields $\bbF$.
In the following theorems, the rank function $\symrank(f_H, k, n)$ is defined over the field $\bbF$.

\begin{theorem}\label{thm:uniform-exponential-rank-boundedness}
For any graph parameter defined by the graph homomorphism $f_H = \hom(\cdot, H)$,
 we have $f_H(K_0) = 1$ and $\symrank (f_H, k, n) \leqslant |V(H)|^k$ for all $k, n \ge 0$. 
\end{theorem}
\begin{proof}
The first claim is obvious, as an empty product is $1$,
and the sum in~(\ref{eqn:hom-def})
 is over the unique empty map $\emptyset$ which is the only possible map
from  the empty set $V(K_0)$.
 For the second claim
  notice that for any $k$-labeled graphs $G_1, \ldots, G_n$ and $\phi \colon [k] \to V(H)$,
\begin{equation}\label{eqn:partial-hom-mult}
\hom_\phi(G_1 \cdots G_n, H) = \hom_\phi(G_1, H) \cdots \hom_\phi(G_n, H).
\end{equation}
When $n = 0$, this equality is $\hom_\phi(U_k, H) = 1$
according to~(\ref{eqn:hom-phi-def}), as an empty product is $1$.

%

By~(\ref{eqn:full-hom-dec}) and~(\ref{eqn:partial-hom-mult}),
for the connection tensor $T(f_H, k, n)$ we have the following decomposition:
\[
T(f_H, k, n) = \sum_{\phi \colon [k] \to V(H)} \alpha_\phi (\hom_\phi(\cdot, H))^{\otimes n}
\]
where each $\hom_\phi(\cdot, H) \in \bbF^{\calPLG[k]}$ and $k, n \ge 0$. 
Let $q = |V(H)|$.
Then each $T(f_H, k, n)$ is a linear combination of $q^k$ tensor $n$-powers
and therefore $\symrank  T(f_H, k, n) \le q^k$ for $k, n \ge 0$.
\end{proof}

The main results of this paper are Theorems~\ref{thm:uniform-exponential-rank-boundedness-converse}
and~\ref{thm:uniform-exponential-rank-boundedness-converse-strengthened}, a converse
to Theorem~\ref{thm:uniform-exponential-rank-boundedness}.
\begin{theorem}\label{thm:uniform-exponential-rank-boundedness-converse}
Let $f$ be a graph parameter for which $f(K_0) = 1$ and there exists a nonnegative integer $q$ such that $\symrank(f, k, n) \le q^k$ for every $k, n \ge 0$. Then there exists a weighted graph $H$ with $|V(H)| \le q$ such that $f = f_H$.
\end{theorem}
More generally,  we have the following stronger theorem.
\begin{theorem}\label{thm:uniform-exponential-rank-boundedness-converse-strengthened}
Let $f$ be a graph parameter for which $f(K_0) = 1$ and there exists a nonnegative integer $q$ such that for every $k \ge 0$ there exists $n \ge 2$ such that $\symrank(f, k, n) \le \min(n - 1, q^k)$. Then there exists a weighted graph $H$ with $|V(H)| \le q$ such that $f = f_H$.
\end{theorem}

Theorem~\ref{thm:uniform-exponential-rank-boundedness-converse-strengthened} implies Theorem~\ref{thm:uniform-exponential-rank-boundedness-converse} 
by choosing a large $n$. Indeed if $\symrank(f, k, n) \le q^k$,
we may choose any $n \ge \max(q^k + 1, 2)$, which is 
$q^k + 1$ unless $q=0$ and $k>0$.

In Section~\ref{sec:proof-of-main-results} we will prove Theorem~\ref{thm:uniform-exponential-rank-boundedness-converse-strengthened}, then Theorem~\ref{thm:uniform-exponential-rank-boundedness-converse} also follows.

%% file: applications-merged.tex
\section{Applications}\label{sec:apps} 


 
%

\subsection{Tensor rank lower bound of certain tensors}\label{subsec:tensor-rank-lower-bound}



We first 
prove 
a lemma about the
rank of the connection tensor for graph matchings.
Let $\calM_{a, b} = \calM_{n; a, b} \in \Sym^n(\bbF^2)$
denote the function $\{0, 1\}^n \rightarrow \bbF$ ($n \ge 0$),
such that on the all-$0$ input $\mathbf 0$ it takes value $a$,
on all inputs of Hamming weight one it takes value $b$,
and on all other inputs it takes value $0$.
This function is denoted by $[a, b, 0, \ldots, 0]$ in the Holant literature.
($\calM_{0; a, b}$ is just a constant $a$.)
We have the following lemma;
the proof
is adapted from the proof of Lemma 5.1 in~\cite{Comon-Golub-Lim-Mourrain-2008}.

\begin{lemma}\label{lem:weighted-matchings-tensor-rank-lower-bound}
If $b \ne 0$ and $n \ge 0$, then $\symrank \calM_{n; a, b} \ge n$.
\end{lemma}
\begin{proof}
For $n = 0$ the lemma is trivial. Let $n \ge 1$.
Clearly $\calM_{n; a, b} \not = 0$, and so  $r = \symrank \calM_{n; a, b} \ge 1$.
Suppose  $r < n$ for a contradiction. 
Then we can write $\calM_{n; a, b} = \sum_{i = 1}^r \lambda_i v_i^{\otimes n}$ 
where  $\lambda_i \in \bbF$, and $v_i = (\alpha_i, \beta_i) \in \bbF^2$
are nonzero and 
pairwise linearly independent.
The decomposition implies that
the linear system $A x = \vec b$ with the extended matrix
\begin{equation}\label{eqn:wm-vand-sys}
\widehat A = [A \mid \vec b] =
\left[
\begin{array}{cccc|c}
\alpha_1^n & \alpha_2^n & \ldots & \alpha_r^n & a \\
\alpha_1^{n - 1} \beta_1 & \alpha_2^{n - 1} \beta_2 & \ldots & \alpha_r^{n - 1} \beta_r & b \\
\alpha_1^{n - 2} \beta_1^2 & \alpha_2^{n - 2} \beta_2^2 & \ldots & \alpha_r^{n - 2} \beta_r^2 & 0 \\
\vdots & \vdots & \ddots & \vdots & \vdots \\
\beta_1^n & \beta_2^n & \ldots & \beta_r^n & 0
\end{array}
\right]
\end{equation}
has a solution $x_i = \lambda_i,\, 1 \le i \le r$.
Note that $\widehat A$ is $(n+1) \times (r+1)$ and
$A$ has only $r$ columns.
We show that $\matrank \,\widehat A = r+1 > \matrank \, A$.
This is a contradiction. We consider the following two cases.

\begin{enumerate}
\item All $\beta_i \ne 0$.
Then, by the pairwise linear independence of $v_i$, the ratios 
 $\alpha_i / \beta_i$ are pairwise distinct.
Then the last $r$ rows of $A$, i.e., rows $n - r + 2$ to $n + 1$
form an $r \times r$ Vandermonde matrix of rank $r$.
Note that $n+1 \ge n - r + 2 > 2$.
By $b \ne 0$, we get an $(r + 1) \times (r + 1)$ submatrix
of  $\widehat A$  of rank $r+1$ by taking row $2$ and the last $r$ rows.

\item Some  $\beta_i = 0$. Without loss of generality we can assume it is $\beta_1$.
Again by the pairwise linear independence of $v_i$, 
 all  other $\beta_i \not = 0$,
and all $\alpha_i / \beta_i$ are pairwise distinct for $2 \le i \le r$
(vacuously true if $r=1$).
Then since $b \ne 0$, the submatrix  of $\widehat A$
formed by taking rows $1, 2$ and the last $r - 1 \ge 0$
rows have rank $r+1$.
\end{enumerate}
\end{proof}


We now show that for an infinite field $\bbF$, 
we can give a tight
 upper bound for $\symrank \calM_{n; a, b}$ where $b \ne 0$  and $n \ge 1$.
The existence of
a decomposition $\calM_{n; a, b} = \sum_{i = 1}^r \lambda_i v_i^{\otimes n}$
where $r \ge 1$, $\lambda_i \in \bbF$ and $v_i = (\alpha_i, \beta_i) \in \bbF^2$
($1 \le i \le r$)
is equivalent to the statement that system~(\ref{eqn:wm-vand-sys})
has a solution $x_i = \lambda_i\, (1 \le i \le r)$.
Note that by Lemma~\ref{lem:weighted-matchings-tensor-rank-lower-bound},
we must have $r \ge n$.

Assume $\bbF$ is infinite. 
We show how to achieve $r = n$ with one exceptional case.
First, we set all $\beta_i = 1$.
By comparing with the Vandermonde determinant
$\det([A \mid \vec t\,])$ as a polynomial in $t$, where the last column is
 $\vec t = (t^n, t^{n - 1}, \ldots, t, 1)^T$,
we have 
\[
\det \widehat A  = \det([A \mid \vec b]) = (-1)^{n} \prod_{1 \le i < j \le n} (\alpha_i - \alpha_j) \cdot \Big(a - b \sum_{i = 1}^n \alpha_i\Big).
\]
To see this equation, note that as a polynomial in $t$ the Vandermonde determinant $\det([A \mid \vec t\,]) = \prod_{1 \le i < j \le n} (\alpha_i - \alpha_j) \prod_{i = 1}^n (\alpha_i - t)= \sum_{i = 0}^n c_i t^i$ for 
some $c_i \in \bbF$ ($0 \le i \le n$).
Then $\det \widehat A = a c_n + b c_{n - 1}$. 

If we set $\alpha_1, \ldots, \alpha_n  \in \bbF$ 
to be pairwise distinct, and $\sum_{i = 1}^n \alpha_i = a/b$,
then $\matrank A = \matrank \hat A = n$.
The (affine) hyperplane ${\mathbf \Pi}: \sum_{i = 1}^n \alpha_i = a/b$
has points away from its intersections with finitely many
hyperplanes $x_i = x_j$ ($i \ne j$), as long as
each of these hyperplanes is distinct from ${\mathbf \Pi}$.
This is trivially true if $n=1$. Let $n \ge 2$.
Under an affine linear transformation 
we may assume ${\mathbf \Pi}$ is the hyperplane $x_n =0$ in $\bbF^n$
and we only need to show $\bbF^{n-1}$ is not the union of finitely many,
say $k$, affine hyperplanes.
Consider the cube $S^{n-1}$ for a large subset $S \subseteq \bbF$.
The union of these  $k$ affine hyperplanes intersecting $S^{n-1}$
 has cardinality at 
most $k|S|^{n-2} < |S|^{n-1}$, for a large $S$.

Each hyperplane  $x_i = x_j$ ($i \ne j$) is distinct from ${\mathbf \Pi}$,
except in one case
\begin{equation}\label{case:exception}
a=0, ~n=2, ~\mbox{and}~ \Char \bbF = 2.
\end{equation}
In this exceptional case, we can easily prove that indeed 
 $\symrank \calM_{2; 0, b} =3$.

We have proved the following. 
\begin{lemma}\label{lem:weighted-matchings-tensor-rank-upper-bound}
If $\bbF$ is infinite, $b \ne 0$ and $n \ge 1$, then $\symrank \calM_{n; a, b} = n$
with one exception (\ref{case:exception}).
In that exceptional case, $\symrank \calM_{2;0, b} = 3$.
\end{lemma}
We remark that for any infinite $\bbF$
not in case (\ref{case:exception}), for $n \ge 2$
we can achieve $\matrank A = \matrank \widehat A = n$ in (\ref{eqn:wm-vand-sys})
by further requiring that all $\alpha_i \not =0$ (in addition to
being pairwise distinct, and all $\beta_i=1$.)
This is simply to avoid the  intersections of ${\mathbf \Pi}$
 with  another finitely many  hyperplanes distinct from ${\mathbf \Pi}$.
Setting $a_i = 1/\alpha_i$, we can set $a_i\in \bbF$ such that
 $\matrank A = \matrank \widehat A = n$ for the following $\widehat A$.
\begin{equation}\label{eqn:wm-vand-sys-for-d-deg-bounded-graphs}
\widehat A = [A \mid \vec b] =
\left[
\begin{array}{cccc|c}
1 & 1  & \ldots  & 1 & a\\
a_1 & a_2 & \ldots & a_n & 1 \\
\vdots & \vdots & \ddots & \vdots & \vdots \\
a_1^{n-1} & a_2^{n-1} & \ldots & a_n^{n-1} & 0\\
a_1^n & a_2^n & \ldots & a_n^n & 0 
\end{array}
\right].
\end{equation}
The linear system $A x = \vec b$ has a solution $(\lambda_1, 
\ldots, \lambda_n)$ implies that
 $\calM_{m; a, b} = \sum_{i = 1}^n \lambda_i v_i^{\otimes m}$,
where $v_i = (1, a_i)$,
for  all $0 \le m \le n$.
(For $m=0$, $\calM_{0; a, 1} = a$ is a constant, and $v_i^{\otimes 0} = 1$.)


\subsection{Perfect matchings~\label{sec:Perfect-Matchings}}

Let  $\mathcal{F}$ be any set of
$\bbF$-valued constraint functions, a.k.a. signatures,
from some finite set $[q]$.
E.g., the  binary {\sc Equality} $(=_2)$ signature
on $(x, y)$ outputs value 1 if $x=y$, and 0 otherwise.
Similarly one can define {\sc All-Distinct} on $[q]$,
and {\sc Exact-One} and {\sc Exact-Two}
 on the Boolean domain $(q=2)$.
An input to a Holant problem $\Holant(\mathcal{F})$ is 
 $\Omega = (G, \pi)$
where  $G = (V,E)$ is  a graph (with possible multiple edges and loops),
and $\pi$ assigns to each $v \in V$ some $f_v \in \mathcal{F}$
of arity $\deg(v)$, and associate its incident edges as input variables
 to $f_v$.
 The output is $\Holant(G; \mathcal{F})
 = \sum_{\sigma} \prod_{v \in V} f_v(\sigma \mid_{E(v)})$, where the sum is  over all edge assignments $\sigma: E \to [q]$,
 $E(v)$ denotes the incident edges of $v$ and $\sigma \mid_{E(v)}$ denotes the restriction of $\sigma$.
Bipartite $\Holant(G;  \mathcal{F} \mid \mathcal{G})$ are
defined on bipartite graphs $G = (U, V, E)$ where vertices in
$U$ and $V$ are assigned signatures from $\mathcal{F}$ and $\mathcal{G}$
respectively.

The graph parameter that counts the number of perfect matchings in a graph,
denoted by \textsc{\#Perfect-Matching} (or $\pmatch$),
is a quintessential Holant problem, corresponding to the {\sc Exact-One} function.
In this subsection we show it 
is not expressible as a GH function over any field.
This was proved in~\cite{Freedman-Lovasz-Schrijver-2007} for GH functions with real edge and positive vertex weights.
 However
that proof does not work for arbitrary fields, 
e.g., for the field of complex numbers $\bbC$,
 or even for real numbers with arbitrary (not necessarily positive) vertex weights. A crucial condition in~\cite{Freedman-Lovasz-Schrijver-2007} is positive semidefiniteness. 
Our main result (Theorems~\ref{thm:uniform-exponential-rank-boundedness}, \ref{thm:uniform-exponential-rank-boundedness-converse} and~\ref{thm:uniform-exponential-rank-boundedness-converse-strengthened})
indicates that the property of being expressible as a GH function
is completely characterized by tensor rank. 

Let $\pmatch(G) = m \cdot 1 \in \bbF$ (the sum of $m$ copies of $1 \in \bbF$) where 
 $m$ is the  number of perfect matchings in $G$.
Obviously, $\pmatch$ is a multiplicative graph parameter with $\pmatch(K_0) = 1$. Next, let $G$ be a $k$-labeled graph, let $X \subseteq [k]$, and let $\pmatch(G, X)$ denote the number of matchings in $G$ (expressed in $\bbF$)
that match all the unlabeled nodes and, for labeled nodes,
exactly the nodes in $X$. 
Then 
for any $k$-labeled graphs $G_1, \ldots, G_n$,
\[
\pmatch(G_1 \cdots G_n) = \sum_{X_1 \sqcup \ldots \sqcup X_n = [k]} \pmatch(G_1, X_1) \cdots \pmatch(G_n, X_n).
\]

This means that $T(\pmatch, k, n)$ is 
the product $N_k^{\otimes n} W_{k, n}$ where $N_k$ has 
infinitely many rows indexed by all $k$-labeled graphs $G$, but only $2^k$ columns indexed by the subsets $X$ of $[k]$, with the entry at $(G, X)$
\[
N_{k; G, X} = \pmatch(G, X),
\]
and $W_{k, n}$ is a symmetric $\underbrace{2^k \times \ldots \times 2^k}_{n \text{ times}}$ tensor (from $\Sym^n(\bbF^{2^k})$), where
\[
W_{k, n; X_1, \ldots, X_n} =
\begin{lcases}
& 1 \quad \text{if } X_1 \sqcup \ldots \sqcup X_n = [k], \\
& 0 \quad \text{otherwise.}
\end{lcases}
\]
For any $k$,  if  $W_{k, n} = \sum_{i=1}^r a_i v_i^{\otimes n}$, then
$T(\pmatch, k, n) = \sum_{i=1}^r a_i (N_{k} v_i)^{\otimes n}$.
Hence 
 $\symrank T(\pmatch, k, n) \le \symrank W_{k, n}$. 
We show that in fact equality holds. Consider the family of $k$-labeled graphs $\{ P_X \}_{X \subseteq [k]}$ of cardinality $2^k$ indexed by the subsets of $[k]$ and defined as follows: each $P_X$ has $|X|$ unlabeled vertices $\{ x_i \}_{i \in X}$ and $k$ labeled vertices $\{ y_i \}_{i = 1}^k$ labeled $1$ to $k$,
 with an edge between $x_i$ and $y_j$ iff $i = j$. It is easy so see that for $X, Y \subseteq [k]$, $N_{k; P_X, Y} = 1$ if $X = Y$ and $0$ otherwise. Then if we consider the subset of rows in $N_k$ corresponding to $\{ P_X \}_{X \subseteq [k]}$ we see that they form the identity matrix $I_{2^k}$ with a suitable
order of rows. Therefore $\symrank W_{k, n} = \symrank \left( I_{2^k}^{\otimes n} W_{k, n} \right) \le \symrank T(\pmatch, k, n)$ and so $\symrank T(\pmatch, k, n) = \symrank W_{k, n}$.

Note that for $k = 1$, $W_{1, n}$ is just the perfect matching
tensor (or the {\sc Exact-One} function on $n$ inputs) $\calM_{0, 1}
 \in \Sym^n(\bbF^2)$ where $n \ge 1$. Applying Lemma \ref{lem:weighted-matchings-tensor-rank-lower-bound} with $a = 0, b = 1$, we get $\symrank W_{1, n} \ge n$ and therefore $\symrank T(\pmatch, 1, n) \ge n$ for $n \ge 1$. Now if $\pmatch$ were expressible as $\hom(\cdot, H)$ for some weighted graph $H$ with $q = |V(H)|$, then by Theorem \ref{thm:uniform-exponential-rank-boundedness}, $\symrank T(\pmatch, k, n) \le q^k$ for $k, n \ge 0$ so that $\symrank T(\pmatch, 1, n) \le q$ for $n \ge 0$. However, as we have just shown $\symrank T(\pmatch, 1, n) \ge n$ for $n \ge 1$ which contradicts the upper bound when $n > q$. Hence $\pmatch$ is not expressible as a graph homomorphism function over any field. We state it as a theorem:
\begin{theorem}\label{thm:Perfect-Matching-not-expressible}
The graph parameter \textsc{\#Perfect-Matching} \emph{($\pmatch$)} is not expressible as a graph homomorphism function over any field.
\end{theorem}

In this proof we have only used simple $k$-labeled graphs that  do
not have edges between the $k$ labeled vertices. The graphs $\{ P_X \}_{X \subseteq [k]}$ clearly have this property, and this property is
preserved under product of 
$k$-labeled graphs. 
It follows that Theorem~\ref{thm:Perfect-Matching-not-expressible}
holds even when $\pmatch$ is restricted to simple graphs. 

We can prove the same inexpressibility results for
other Holant problems, such as weighted matchings,
proper edge colorings, and
vertex disjoint cycle covers.
We will also discuss bounded degree cases of weighted matchings.

\subsection{Weighted matchings~\label{sec:Weighted-Matchings}}

We prove that the problem \textsc{\#Weighted-Matching$_a$} defined by
$\wmatch_a(G) = \Holant(G; \{ \calM_{n; a, 1} \}_{n \ge 0})$ 
is not expressible as a GH function over any field,
for any $a$.

Clearly, $\wmatch_a$ is a multiplicative graph parameter with $\wmatch_a(K_0) = 1$. If  $a = 0$ then  $\wmatch_a$ is just \textsc{\#Perfect-Matching} ($\pmatch$). Counting all  matchings
is  $\wmatch_a$ for $a=1$.

Let $G$ be a $k$-labeled graph,  $X \subseteq [k]$,
and let $\wmatch_a(G, X)$ denote the partial $\Holant$ 
sum in $\wmatch_a(G)$
over all $\{ 0, 1 \}$-edge assignments of $G$ such that
within $[k]$, those in $X$ have  exactly one
 incident edge assigned $1$ and all nodes in $[k] \setminus X$ have no
incident edges assigned $1$.  
Then we have for any $k$-labeled graphs $G_1, \ldots, G_n$,
\[
\wmatch_a(G_1 \cdots G_n) = \sum_{X_1 \sqcup \ldots \sqcup X_n \subseteq [k]} a^{k - |X_1 \sqcup \ldots \sqcup X_n|} \wmatch_a(G_1, X_1) \cdots \wmatch_a(G_n, X_n).
\]
This means that $T(\wmatch_a, k, n)$ is
the product  $N_{k; a}^{\otimes n} W_{k, n; a}$,  where $N_{k; a}$ has
infinitely many rows indexed by all $k$-labeled graphs $G$, 
and $2^k$ columns indexed by $X \subseteq [k]$, with the entry at $(G, X)$
\[
N_{k; a; G, X} = \wmatch_a(G, X),
\]
and $W_{k, n; a}$ is a symmetric $\underbrace{2^k \times \ldots \times 2^k}_{n \text{ times}}$ tensor (from $\Sym^n(\bbF^{2^k})$), where
\[
W_{k, n; a; X_1, \ldots, X_n} =
\begin{lcases}
& a^{k - |X_1 \sqcup \ldots \sqcup X_n|} \quad \text{if } X_1 \sqcup \ldots \sqcup X_n \subseteq [k], \\
& 0 \quad \text{otherwise.}
\end{lcases}
\]
Hence 
 $\symrank T(\wmatch_a, k, n) \le \symrank W_{k, n; a}$. We show that in fact equality holds. Consider the same family of $k$-labeled graphs
 $\{ P_X \}_{X \subseteq [k]}$ defined in the previous subsection.
It is easy to see that for $X, Y \subseteq [k]$, $N_{k; a; P_X, Y} = a^{|X| - |Y|}$ if $Y \subseteq X$ and $0$ otherwise. Here by convention, $a^0 = 1$ even if $a = 0$. Consider the rows in $N_{k; a}$ corresponding to $\{ P_X \}_{X \subseteq [k]}$. 
They form the nonsingular matrix $\trans{1}{0}{a}{1}^{\otimes n}$ 
if the rows and columns are ordered lexicographically for $X, Y \subseteq[k]$.
Therefore $\symrank W_{k, n; a} = \symrank 
\left(\trans{1}{0}{a}{1}^{\otimes n} W_{k, n; a} \right) \le \symrank T(\wmatch_a, k, n)$ and so $\symrank T(\wmatch_a, k, n) = \symrank W_{k, n; a}$.

Note that for $k = 1$, $W_{1, n; a} = \calM_{n;a, 1} = [a, 1, 0, \ldots 0] \in \Sym^n(\bbF^2)$ where $n \ge 1$. Applying Lemma~\ref{lem:weighted-matchings-tensor-rank-lower-bound} with $b = 1$, we get $\symrank W_{1, n; a} \ge n$ and therefore $\symrank T(\wmatch_a, 1, n) \ge n$ for $n \ge 1$. Now if $\wmatch_a$ were expressible as $\hom(\cdot, H)$ for some weighted graph $H$ with $q = |V(H)|$, then by Theorem~\ref{thm:uniform-exponential-rank-boundedness}, $\symrank T(\wmatch_a, k, n) \le q^k$ for $k, n \ge 0$ so that $\symrank T(\wmatch_a, 1, n) \le q$ for $n \ge 0$. 
This contradicts  $\symrank T(\wmatch_a, 1, n) \ge n$ when $n > q$.
Hence $\wmatch_a$ is not expressible as a graph homomorphism function over any field. 

By the same remark for
Theorem~\ref{thm:Perfect-Matching-not-expressible}, the proof for
 Theorem~\ref{thm:weighted-Matching-not-expressible-simple}
 carries over to simple graphs.

\begin{theorem}\label{thm:weighted-Matching-not-expressible-simple}
The graph parameter \textsc{\#Weighted-Matching} \emph{($\wmatch_a$)} where $a \in \bbF$ as a function defined on simple graphs is not expressible as a graph homomorphism function over any field $\bbF$.
\end{theorem}

\input{bounded-degree-graphs}
\subsection{Proper edge \texorpdfstring{$d$}{d}-colorings}

Next we show that the graph parameter  \#$d$-\textsc{Edge-Coloring}
($\edgecol_d$) is not expressible as
a GH function over any field of characteristic $0$. Given a graph $G$,
$\edgecol_d(G)$ counts the number of proper edge $d$-colorings in a graph,
 where $d \ge 1$ is the number of colors available.

Clearly, $\edgecol_d$ is a multiplicative graph parameter with $\edgecol(K_0) = 1$. Since $\Char \bbF = 0$, $\bbF$ is infinite. 
Consider $K_1$ as a $1$-labeled graph and the star graph $S_d$
with one internal node labeled by $1$ and $d$ unlabeled external nodes
all connected to node 1, where $d \ge 1$. 
Consider the  connection tensor $T(\edgecol_d, k, n)$ 
restricted to $\{ K_1, S_d \}^n$. For $(G_1, \ldots, G_n) \in \{ K_1, S_d \}^n$,
 if more than one $G_i = S_d$ then the product $G_1 \cdots G_n$ has no
proper edge $d$-coloring because the labeled vertex
has degree $\ge 2d > d$. Then
 it is easy to see that the connection tensor $T(\edgecol_d, k, n)$ restricted to $\{ K_1, S_d \}^n$ has the
 form $\calM_{1, d!} = [1, d!, 0, \ldots, 0] \in \Sym^n(\bbF^2)$,
and $d! \ne 0$ in $\bbF$ as $\Char \bbF = 0$.
Therefore by Lemma~\ref{lem:weighted-matchings-tensor-rank-lower-bound}, $\symrank \calM_{1, d!} \ge n$ for $n \ge 1$. Hence $\symrank T(\edgecol_d, 1, n) \ge n$ for $n \ge 1$.

Now if $\edgecol_d$ were expressible as $\hom(\cdot, H)$ for some weighted graph $H$ with $q = |V(H)|$, then by Theorem \ref{thm:uniform-exponential-rank-boundedness},
 $\symrank T(\edgecol_d, 1, n) \le q$ for $n \ge 0$. This contradicts the upper bound when $n > q$. Hence $\edgecol_d$ is not expressible as a graph homomorphism function over any field $\bbF$ of $\Char \bbF = 0$.

By the same remark for
Theorem~\ref{thm:Perfect-Matching-not-expressible},
the proof for
 Theorem~\ref{thm:Edge-Coloring-not-expressible-simple}
 carries over to  simple graphs.

\begin{theorem}\label{thm:Edge-Coloring-not-expressible-simple}
The graph parameter \#$d$-\textsc{Edge-Coloring} \emph{($\edgecol_d$)} with $d \ge 1$ as a function defined on simple graphs is not expressible as a graph homomorphism function over any field of characteristic $0$.
\end{theorem}

\subsection{Vertex-disjoint cycle covers}

We show that the graph parameter $\vdccover$
(\#\textsc{Vertex-Disjoint-Cycle-Cover})
 which counts the number of vertex disjoint cycle covers in a graph
is not expressible as a GH function over an arbitrary field.
In a multigraph without loop a cycle is
 a vertex disjoint closed path of length at least $2$.
The graph parameter $\vdccover(G) = m \cdot 1 \in {\mathbb F}$, where $m$ is
the number of edge subsets $E'$ that 
form  a vertex disjoint set of cycles that cover all vertices.

Clearly, $\vdccover$ is a multiplicative graph parameter with $\vdccover(K_0) = 1$. Next, consider $K_1$ and $K_3$ as $1$-labeled graphs.
Note that $K_3$ is a cycle of 3 vertices.
 It is easy to see that the connection tensor $T(\vdccover, k, n)$ restricted to $\{ K_1, K_3 \}^n$ has the form $\calM_{0, 1} = [0, 1, 0, \ldots, 0] \in \Sym^n(\bbF^2)$ and therefore by Lemma~\ref{lem:weighted-matchings-tensor-rank-lower-bound}, $\symrank \calM_{0, 1} \ge n$ for $n \ge 1$. Hence $\symrank T(\vdccover, 1, n) \ge n$ for $n \ge 1$.

Now if $\vdccover$ were expressible as $\hom(\cdot, H)$ for some weighted graph $H$ with $q = |V(H)|$, then by Theorem \ref{thm:uniform-exponential-rank-boundedness},  $\symrank T(\vdccover, 1, n) \le q$ for $n \ge 0$. This contradicts the upper bound when $n > q$. Hence $\vdccover$ is not expressible as a 
GH function over any field. 

By the same remark for
Theorem~\ref{thm:Perfect-Matching-not-expressible},
the proof for
 Theorem~\ref{thm:Vertex-Disjoint-Cycle-Cover-not-expressible-simple}
 carries over to  simple graphs.

\begin{theorem}\label{thm:Vertex-Disjoint-Cycle-Cover-not-expressible-simple}
The graph parameter \#\textsc{Vertex-Disjoint-Cycle-Cover} \emph{($\vdccover$)} as a function defined on simple graphs is not expressible as a graph homomorphism function over any field.
\end{theorem}

%% file: bounded-degree-graphs.tex
\subsection{Bounded degree graphs}\label{sec:bonded-deg-graphs}

Fix any $d \ge 2$.
A degree-$d$ bounded graph is a graph with maximum degree at most $d$.
In this subsection, we investigate the expressibility of the graph parameter \textsc{\#Weighted-Matching$_a$} ($\wmatch_a$) 
as a GH function on bounded degree graphs.  More precisely, we are interested when $\wmatch_a$ is expressible as a $\hom(\cdot, H)$ with $|V(H)| = q$ on degree-$d$ bounded graphs. 
For convenience, we temporarily allow vertex weights to be $0$,
and it will be addressed later.


Given a graph $G$ (possibly with multiple edges but no loops), 
let $G'$ be the vertex-edge incidence graph of $G$.
The vertex set $V(G')$
consists of the original vertices from $V(G)$ on the LHS 
and the edges $E(G)$
on the RHS. 
Let $H$ be the weighted graph specified by 
vertex weights $(\alpha_1, \ldots, \alpha_q) \in \bbF^q$ and a 
symmetric matrix $B = (\beta_{i j}) \in \bbF^{q \times q}$ for edge
weights, then for any $G$
\begin{equation}\label{eqn:hom-to-bip-holant}
\hom(G, H) = \Holant(G'; \{ \sum_{i = 1}^q \alpha_i e_{q, i}^{\otimes n} \}_{n \ge 0} \mid B),
\end{equation}
where $\{ e_{q, i} \}_{i = 1}^q \in \bbF^q$ 
has a single 1 at the $i$th position and 0 elsewhere.
Here $\hom(G, H)$ is expressed in (\ref{eqn:hom-to-bip-holant})
 as a domain-$q$ Holant sum on $G'$: any LHS vertex of $G'$ of degree $n$ is assigned the signature $\sum_{i = 1}^q \alpha_i e_{q, i}^{\otimes n}$ (which takes value $\alpha_i$ if all
incident edges have value $i \in [q]$, and 0 otherwise),
and any RHS vertex of $G'$ (an edge of $G$)
is assigned the symmetric binary signature specified by $B$. 

First, we show that for any $a \in \bbF$, if
 $\wmatch_a$ is 
expressible as $\hom(\cdot, H)$ with $|V(H)| = q$ on degree-$d$ 
bounded \emph{simple} graphs over any field $\bbF$, then $q \ge d$.
 Recall the proof
from Section~\ref{sec:Weighted-Matchings}. This time we restrict the connection tensor $T(\wmatch_a, 1, d)$ (for $k = 1$) to the $1$-labeled graphs
$P_\emptyset$ and $P_{[1]}$, which are just $K_1$ and $K_2$ without
the label. 
The product of $d$ labeled graphs from $\{P_\emptyset, P_{[1]}\}$
having $\ell$ copies of $P_{[1]}$
is the star graph $S_\ell$ with  one internal node labeled by $1$ and 
$\ell$ external unlabeled nodes. 
All these are degree-$d$ bounded simple graphs. 
Note that $T(\wmatch_a, 1, d)_{|\{ P_\emptyset, P_{[1]} \}^d} 
= \trans{1}{0}{a}{1}^{\otimes d} W_{1, d; a}$. Therefore $\symrank \left( T(\wmatch_a, 1, d)_{|\{ P_\emptyset, P_{[1]} \}^d} \right) = \symrank \left(\trans{1}{0}{a}{1}^{\otimes d} W_{1, d; a}\right) = \symrank W_{1, d; a} \ge d$, the last step is by Lemma~\ref{lem:weighted-matchings-tensor-rank-lower-bound}. On the other hand, if $\wmatch_a$ is expressible as $\hom(\cdot, H)$ with $|V(H)| = q$ on degree-$d$ bounded \emph{simple} graphs, then arguing similarly to the proof of Theorem~\ref{thm:uniform-exponential-rank-boundedness} but restricting the domain of the arguments $G_i, \, 1 \le i \le d$ in (\ref{eqn:partial-hom-mult})
to $\{P_\emptyset, P_{[1]} \}$, we have 
$\symrank \left( T(\wmatch_a, 1, d)_{|\{ P_\emptyset, P_{[1]} \}^d} \right) \le q$ so $d \le q$.
Then clearly this bound also holds if we do not allow $0$-weighted vertices.

Now let $\bbF$ be infinite.
By the remark after 
Lemma~\ref{lem:weighted-matchings-tensor-rank-upper-bound},
 if we are not
in the exceptional case (\ref{case:exception}) (we put $n = d$), then
for some
$a_i, \alpha_i \in \bbF$  we have
 $\calM_{m; a, 1} = \sum_{i = 1}^d \alpha_i (1, a_i)^{\otimes m}$,
for every $0 \le m \le d$.
 Let $T \in \bbF^{d \times 2}$ be the matrix 
whose $1$st and $2$nd columns 
are $(1, 1, \ldots, 1)^{\tt T}$ and $(a_1, a_2, \ldots, a_d)^{\tt T}$
respectively. Then $e_{d, i} T = (1, a_i)$.
Define the symmetric matrix $B = (\beta_{i j}) =
 T  T^{\tt T} \in \bbF^{d \times d}$. Now let $H$ be a weighted graph on $d$ vertices specified by $(\alpha_1, \ldots, \alpha_d) \in \bbF^d$ and $B = (\beta_{i j}) \in \bbF^{d \times d}$. Then for any degree-$d$ bounded graph $G$ we have the following equality chain:
\begingroup
\allowdisplaybreaks
\begin{align*}
\wmatch_a(G) &= \Holant(G; \{ \calM_{m; a, 1} \}_{0 \le m \le d}) = \Holant(G'; \{ \calM_{m; a, 1} \}_{0 \le m \le d} \mid (=_2) ) \\
&= \Holant(G'; \{ \sum_{i = 1}^d \alpha_i (1, a_i)^{\otimes m} \}_{0 \le m \le d} \mid (=_2) ) = \Holant(G'; \{  \sum_{i = 1}^d \alpha_i (e_{d, i} T )^{\otimes m} \}_{0 \le m \le d} \mid (=_2) ) \\
&= \Holant(G'; \{ \sum_{i = 1}^d \alpha_i e_{d, i}^{\otimes m} T^{\otimes m} \}_{0 \le m \le d} \mid (=_2) ) = \Holant(G'; \{ \sum_{i = 1}^d \alpha_i e_{d, i}^{\otimes m} \}_{0 \le m \le d} \mid T^{\otimes 2} (=_2) ).
\end{align*}
\endgroup
where the last equation moving $T^{\otimes n}$ from the left-hand side of
the Holant problem to $T^{\otimes 2}$ in the right-hand side,
is called a holographic transformation~\cite{Valiant-2004, cai-chen-book} (the argument works for arbitrary fields).
This follows from the associativity of the operation of tensor 
contraction. 
The {\sc Equality} function  $(=_2)$ is transformed
to $T^{\otimes 2} (=_2)$, which has the matrix form $T T^{\tt T}
= B$. Hence this is precisely the function $\hom(G, H)$.
We have temporarily allowed $0$-weighted vertices;
but in fact by the lower bound $q \ge d$ no $0$-weighted
vertex exists, since otherwise by removing  $0$-weighted vertices we would 
have $\wmatch_a(\cdot) = \hom(\cdot, H')$ with fewer vertices.

The inexpressibility with $|V(H)| = 2$ for the exceptional case
(\ref{case:exception}) holds even for simple graphs, by considering
paths of $0$, $1$ or $2$ edges.
Also, in this case it can be easily shown that
$\wmatch_a = \pmatch$ is expressible as $\hom(\cdot, H)$ where $|V(H)| = 3$:
this can be done similarly to the expressibility proof above via a holographic transformation
(then $H$ cannot have $0$-weighted vertices).


%

This proves Theorem~\ref{thm:wm-as-GH-for-deg-bounded-graphs}.

\begin{theorem}\label{thm:wm-as-GH-for-deg-bounded-graphs}
Let $\bbF$ be a field and $d \ge 2$.
Then for the graph parameter \textsc{\#Weighted-Matching$_a$} \emph{($\wmatch_a$)} where $a \in \bbF$
as a function defined on degree-$d$ bounded graphs the following hold:
\begin{enumerate}
\item $\wmatch_a$ is not expressible as $\hom(\cdot, H)$ with $|V(H)| < d$ even on degree-$d$ bounded \emph{simple} graphs.
\item If $\bbF$ is infinite, then $\wmatch_a$ is expressible as $\hom(\cdot, H)$ with $|V(H)| = d$, 
with one exception
(\ref{case:exception})
in which case the minimal value for $|V(H)|$ is $3$.
\end{enumerate}
\end{theorem}
Note that  \textsc{\#Perfect-Matching} \emph{($\pmatch$)} is just
the special case  $a = 0$. Hence Theorem~\ref{thm:wm-as-GH-for-deg-bounded-graphs} also holds for \emph{$\pmatch$}.

%% file: main-theorem-proof.tex
\section{Proof of Main Theorem}\label{sec:proof-of-main-results}

For now, we do not make any assumptions on the graph parameter $f$;
we will introduce more assumptions as needed to prove the desired statements.
When we speak of submonoids, subrings and subalgebras we require
that the multiplicative identity coincide with that of the larger structure.
When a subset with the induced operations forms a monoid, ring or algebra
we will simply say that it is respectively
a monoid, ring or algebra in the larger structure. 
We \emph{allow} zero algebras and rings, in which $0 = 1$. 
Statements about such structures can be easily checked. 
A function of arity zero is a scalar.
We identify a (labeled) graph with its (labeled) graph isomorphism class.

\subsection{The monoid and algebra of graphs}\label{subsec:monoid-and-algebra-of-graphs}

For every finite set $S \subseteq \posints$, we denote
by $U_S$ the graph with $|S|$ nodes labeled by $S$ and no edges. 
Note that $U_\emptyset = K_0$ is the empty graph.

We put all $k$-labeled graphs into a single structure as follows.
By a \textit{partially labeled graph} we mean a finite graph
in which some of the nodes are labeled by distinct positive integers.
(All label sets are finite.)
Two partially labeled graphs are \textit{isomorphic}
if there is an isomorphism between them preserving all labels.
For two partially labeled graphs $G_1$ and $G_2$,
let $G_1 G_2$ denote the partially labeled graph
obtained by taking the disjoint union of $G_1$ and $G_2$,
and identifying the nodes with the same label;
the union of the label sets becomes the labels of $G_1 G_2$.
This way we obtain a commutative monoid $\calPLG$
consisting of all isomorphism classes of finite partially labeled graphs
with the empty graph $U_\emptyset$ being the identity~\footnote{In \cite{Freedman-Lovasz-Schrijver-2007},
the word semigroup instead of monoid is used.
A monoid is a semigroup with identity,
and all semigroups in~\cite{Freedman-Lovasz-Schrijver-2007}
have or assume to have identity.
Thus, our use of the term monoid is consistent with that of~\cite{Freedman-Lovasz-Schrijver-2007}.}.
For every finite set $S \subseteq \posints$,
we call a partially labeled graph $S$\textit{-labeled},
if its labels form the set $S$.
We call a partially labeled graph ${\subseteq} S$\textit{-labeled},
if its labels form a subset of $S$.
We define $\calPLG(S)$ and $\calPLG_\subseteq(S)$
  to be the subsets of $\calPLG$
consisting of all isomorphism classes of $S$-labeled
and ${\subseteq} S$-labeled graphs, respectively.
Clearly $\calPLG(S) \subseteq \calPLG_\subseteq(S)$.
Then both $\calPLG(S)$ and $\calPLG_\subseteq(S)$ are commutative monoids in $\calPLG$.
$\calPLG_\subseteq(S)$ is a submonoid of $\calPLG$
with the same identity $U_\emptyset$,
while  $\calPLG(S)$ is a submonoid of $\calPLG$ iff $S = \emptyset$,
as the identity in $\calPLG(S)$ is $U_S$.


%

Let $\mathcal G$ denote the
 monoid algebra $\bbF \calPLG$ consisting of all finite formal
linear combinations in $\calPLG$ with coefficients from $\bbF$;
they are called (partially labeled, $\bbF$-)\textit{quantum graphs}.
Restricting the labels to precisely $S$ or to subsets of $S$, 
we have $\bbF \calPLG(S)$ or $\bbF \calPLG_\subseteq(S)$,
 the algebras of  $S$-labeled or  $\subseteq S$-labeled 
\textit{quantum graphs},
denoted  by  $\calG(S)$ or $\calG_\subseteq(S)$, respectively.
$\calG(S)$ is  
an algebra inside $\calG$ with $U_S$ being the multiplicative identity,
and $\calG_\subseteq(S)$ is a subalgebra of $\calG$.
The empty sum is the additive identity in all.

Because many definitions, notations and statements
for $\calPLG(S), \calG(S)$ and $\calPLG_\subseteq(S), \calG_\subseteq(S)$ appear similar, 
we will often commingle them to minimize repetitions,
e.g., we use $\calG_{(\subseteq)}(S)$ to denote
either $\calG(S)$ or $\calG_\subseteq(S)$ (and the statements are asserted for both).

We can extend $f$ to a linear map on $\calG$, and
define an $n$-fold multilinear form,  where $n \ge 1$,
\[
\langle x_1, \ldots, x_n \rangle_{(n)} = f(x_1 \cdots x_n),~~~~~~\mbox{for}~~
x_1, \ldots, x_n \in  \calG.
\]
It is symmetric because $\calG$ is commutative.
Note that if we restrict each argument to $\calG[k]$
and then write it with respect to the basis $\calPLG[k]$ of $\calG[k]$,
we get precisely the connection tensor (array) $T(f, k, n)$.

Let
\[
\calK = \{ x \in \calG \mid \forall y \in \calG, \, f(x y) = \langle x, y \rangle = 0 \}
\]
be the annihilator of $\calG$.
Clearly, $\calK$ is an ideal in $\calG$,
so we can form the quotient algebra $\hat \calG = \calG / \calK$
which is commutative as well.
We denote its identity by $u_\emptyset = U_\emptyset + \calK$.
More generally, we denote $u_S = U_S + \calK$
 for any finite subset $S \subseteq \posints$.
If $x \in \calK$, then $f(x) = f(x U_\emptyset) = 0$
and so $f$ can also be considered as a linear map on $\hat \calG$
by $f(x + \calK) = f(x) + \calK$ for $x \in \calG$.
For a partially labeled graph $G$ we 
denote by $\hat G = G + \calK$ the corresponding element of $\hat \calG$.
More generally, we write $\hat x = x + \calK$ for any $x \in \calG$. 
Since $\calK$ is an ideal in $\calG$, the form
$\langle \cdot, \ldots, \cdot \rangle_{(n)}$ on $\calG$ induces
an $n$-fold multilinear symmetric form on $\hat \calG$,  where $n \ge 1$, 
\begin{equation}\label{eqn:f-on-hat-calG}
\langle x_1, \ldots, x_n \rangle_{(n)} = f(x_1 \cdots x_n),~~~~~~\mbox{for}~~
x_1, \ldots, x_n \in \hat \calG.
\end{equation}

%

We can also define
\[
\hat \calG_{(\subseteq)}(S) = (\calG_{(\subseteq)}(S) + \calK) / \calK = \{ x + \calK \mid x \in \calG_{(\subseteq)}(S) + \calK \}
= \{ x + \calK \mid x \in \calG_{(\subseteq)}(S)\}.
\]
It is easy to see that $\hat \calG_\subseteq(S)$ is a subalgebra of $\hat \calG$
with the same identity $u_\emptyset = U_\emptyset + \calK$,
and $\hat \calG(S)$ is an algebra inside $\hat \calG$
with the identity $u_S = U_S + \calK$.~\footnote{In contrast to \cite{Freedman-Lovasz-Schrijver-2007}
we cannot in general normalize $f$ to make all elements $u_S$
the same in the quotient algebra $\hat \calG$,
 for various finite $S \subseteq \posints$.
This is because in our more general setting, it is possible $f(K_1) = 0$,
in which case the normalization step from \cite{Freedman-Lovasz-Schrijver-2007} fails.
For graph parameters expressible as a graph homomorphism function, this corresponds to the case when all vertex weights sum to $0$.}


If $S, T \subseteq \posints$ are finite subsets,
then $\calPLG_{(\subseteq)}(S) \cdot \calPLG_{(\subseteq)}(T) \subseteq \calPLG_{(\subseteq)}(S \cup T)$
so by linearity we get $\calG_{(\subseteq)}(S) \, \calG_{(\subseteq)}(T) \subseteq \calG_{(\subseteq)}(S \cup T)$
and so, going to the quotients, we have $\hat \calG_{(\subseteq)}(S) \hat \calG_{(\subseteq)}(T) \subseteq \hat \calG_{(\subseteq)}(S \cup T)$.
Also note that for a finite $S \subseteq \posints$,
we have $\calPLG(S) \subseteq \calPLG_\subseteq(S)$
so by linearity $\calG(S) \subseteq \calG_\subseteq(S)$
and then by going to the quotients we obtain $\hat \calG(S) \subseteq \hat \calG_\subseteq(S)$.

Since
$\calG_{(\subseteq)}(S) \cap \calK$ is an ideal in $\calG_{(\subseteq)}(S)$,
 we can also form 
another quotient algebra 
\[
\tilde \calG_{(\subseteq)}(S) = \calG_{(\subseteq)}(S) / (\calG_{(\subseteq)}(S) \cap \calK).
\] 
%
%
We have the following
 canonical isomorphisms between $\tilde \calG_{(\subseteq)}(S)$
 and $\hat \calG_{(\subseteq)}(S)$.
\begin{claim}\label{lem:calG(subseteq)(S)/calK(subseteq)S-canonical-isomorphism} 
Let $S \subseteq \posints$ be finite. Then $\tilde \calG_{(\subseteq)}(S) \cong \hat \calG_{(\subseteq)}(S)$ as algebras via $x + \calG_{(\subseteq)}(S) \cap \calK \mapsto x + \calK,\, x \in \calG_{(\subseteq)}(S)$.
\end{claim}
\begin{proof}
It follows from
the 
Second Isomorphism Theorem for algebraic structures
(see~\cite{Cohn-book} p.~8).
%
%
\end{proof}

For finite $S \subseteq \posints$,
it is convenient to treat the algebras $\tilde \calG_{(\subseteq)}(S)$ and $\hat \calG_{(\subseteq)}(S)$ as separate objects
despite this isomorphism in Claim~\ref{lem:calG(subseteq)(S)/calK(subseteq)S-canonical-isomorphism}.
As it will be seen later, the algebras $\tilde \calG(S)$ with $S = [k]$, where $k \ge 0$, are naturally associated with the $k$th connection tensors $T(f, k, n), n \ge 0$. 
Later we will need to work with various finite $S \subseteq \posints$ simultaneously and need an ambient algebra in which dependencies between elements can be established.
The algebras $\tilde \calG(S)$ do not naturally possess this property as they are the quotients of the algebras $\calG(S)$ which have no common element except $0$.
However, the fact that $\hat \calG(S) \subseteq \hat \calG$ for any finite $S \subseteq \posints$
will allow us to establish dependencies between their elements.
In other words, $\hat \calG$ will serve as the ambient algebra in which further derivations will take place.
Next, the $\subseteq$-definitions will be needed to define a projection $\hat \pi_S \colon \calG \to \calG_\subseteq(S)$ (see Claims~\ref{lem:projection-and-preimage-of-calK-ideals} and~\ref{lem:quotion-decomposition-through-projeciton}). This projection will be used later in the proof. 
%

We say that elements $x, y \in \calG \,(\text{or } \hat \calG)$
are orthogonal (with respect to $f$),
if $f(x y) = 0$ and denote it by $x \perp y$.
For a subset $A \subseteq \calG \,(\text{or } \hat \calG)$, 
denote by $A^\perp = \{ x \in \calG \,(\text{or } \hat \calG) \mid \forall y \in A, \, x \perp y\}$
 the set of those elements  in $\calG \,(\text{or } \hat \calG)$ orthogonal to all elements in $A$.
Next, we say that subsets $A, B \subseteq \calG \,(\text{or } \hat \calG)$
are orthogonal (with respect to $f$),
if $x \perp y$ for all $x \in A$ and $y \in B$.
Similarly, we can talk about an element of $\calG \,(\text{or } \hat \calG)$
being orthogonal to a subset of $\calG \,(\text{or } \hat \calG)$ and vice versa.
Note that the notion of orthogonality is symmetric
since all the multiplication operations considered are commutative.
From the definition, we have $\calK = \calG^\perp$.
Next, denote (commingling the notations $\calK_S$ and $\calK_{\subseteq S}$)
\[
\calK_{(\subseteq) S} = \{ x \in \calG_{(\subseteq)} (S) \mid \forall y \in \calG_{(\subseteq)}(S), \, x \perp y \} = \calG_{(\subseteq)}(S) \cap (\calG_{(\subseteq)}(S))^\perp.
\]
Clearly, $\calK_{(\subseteq) S}$ is an ideal in $\calG_{(\subseteq)}(S)$,
so we can form yet another quotient algebra $\calG_{(\subseteq)}(S) / \calK_{(\subseteq) S}$. 

Next, we define an orthogonal projection
from $\hat \calG$ to the subalgebra $\hat \calG_\subseteq(S)$.
We will show how to do it in a series of lemmas.
Let $S \subseteq \posints$ be finite.
For every partial labeled graph $G$,
let $G_S$ denote the ${\subseteq} S$-labeled graph obtained by deleting
the labels not in $S$ from the vertices of $G$ (unlabeling such vertices).
Extending this map by linearity,
we get a linear map $\pi_S \colon \calG \to \calG_\subseteq(S)$.
Note that $(\pi_S)_{| \calG_\subseteq(S)} = \id_{|\calG_\subseteq(S)}$.
In particular, $\pi_S \colon \calG \to \calG_\subseteq(S)$ is surjective.

\begin{claim}\label{lem:projection} 
Let $S \subseteq \posints$ be finite.
If $x \in \calG$ and $y \in \calG_\subseteq(S)$, then
\[
f(x y) = f(\pi_S(x) y).
\]
\end{claim}
\begin{proof}
For every $G_1 \in \calPLG$ and $G_2 \in \calPLG_\subseteq(S)$,
the graphs $G_1 G_2$ and $\pi_S(G_1) G_2$ are isomorphic as unlabeled graphs.
Hence $f(G_1 G_2) = f(\pi_S(G_1) G_2)$ as $f$ ignores labels.
The claim follows by linearity.
\end{proof}

\begin{claim}\label{cor:projection-part-orthogonality} 
Let $S \subseteq \posints$ be finite.
If $x \in \calG$, then $x - \pi_S(x) \in (\calG_\subseteq(S))^\perp$.
\end{claim}
\begin{proof}
Fix any $y \in \calG_\subseteq(S)$. By Claim~\ref{lem:projection},
$f(x y) = f(\pi_S(x) y)$ so $f((x - \pi_S(x)) y) = 0$.
Thus $x - \pi_S(x) \in (\calG_\subseteq(S))^\perp$.
\end{proof}

So for any $x \in \calG$, we can write $x = \pi_S(x) + (x - \pi_S(x))$
where $\pi_S(x) \in \calG_\subseteq(S)$, and $x - \pi_S(x) \in (\calG_\subseteq(S))^\perp$.
This gives a decomposition $\calG = \calG_\subseteq(S) + (\calG_\subseteq(S))^\perp$.
To get a direct sum decomposition,
we need to pass to the quotient algebra.
But to do so properly we need some more properties.

\begin{claim}\label{lem:calKS} 
Let $S \subseteq \posints$ be finite.
Then $\calK_S = \calG(S) \cap \calK$.
\end{claim}
\begin{proof}
Clearly,
$\calG(S) \cap \calK = \calG(S) \cap \calG^\perp \subseteq
 \calG(S) \cap (\calG(S))^\perp = \calK_S$,
so we only need to prove the reverse inclusion. 
Let $x \in \calK_S = \calG(S) \cap (\calG(S))^\perp$.
Take any $y \in \calG$. Then
\[
f(x y) \stackrel{(1)}{=} f(x \pi_S(y)) \stackrel{(2)}{=} f(x U_S \pi_S(y)) \stackrel{(3)}{=} 0.
\]
Here step $(1)$ uses Claim~\ref{lem:projection} 
as $x \in \calG(S) \subseteq \calG_{\subseteq}(S)$;
$(2)$ is true because $x = x U_S$ for $x \in \calG(S)$;
$(3)$ is true as $\pi_S(y) \in \calG_\subseteq(S)$
so $U_S \pi_S(y) \in \calG(S)$, and as $x \in (\calG(S))^\perp$.
Then $x \in \calK$ so $x \in \calG(S) \cap \calK$,
implying $\calK_S \subseteq \calG(S) \cap \calK$.
\end{proof}

\begin{claim}\label{lem:calKsubseteqS}
Let $S \subseteq \posints$ be finite.
Then $\calK_{{\subseteq}S} = \calG_\subseteq(S) \cap \calK$.
\end{claim}
\begin{proof}
Clearly,
$\calG_\subseteq(S) \cap \calK = \calG_\subseteq(S) \cap \calG^\perp
\subseteq  \calG_\subseteq(S) \cap (\calG_\subseteq(S))^\perp = 
\calK_{{\subseteq}S}$,
so we only need to prove the reverse inclusion. 
Let $x \in \calK_{{\subseteq}S} = \calG_\subseteq(S) \cap (\calG_\subseteq(S))^\perp$.
Take any $y \in \calG$. Then
\[
f(x y) \stackrel{(1)}{=} f(x \pi_S(y)) \stackrel{(2)}{=} 0.
\]
Here step $(1)$ uses Claim~\ref{lem:projection};
$(2)$ is true as $\pi_S(y) \in \calG_\subseteq(S)$
and $x \in (\calG_\subseteq(S))^\perp$.
Then $x \in \calK$ so $x \in \calG_\subseteq(S) \cap \calK$,
implying $\calK_{\subseteq S} \subseteq \calG_\subseteq(S) \cap \cal K$.
\end{proof}

It follows from Claims~\ref{lem:calKS}~and~\ref{lem:calKsubseteqS} that
$\calG_{(\subseteq)}(S) / \calK_{(\subseteq) S} = \calG_{(\subseteq)}(S) / (\calG_{(\subseteq)}(S) \cap \calK) = \tilde \calG_{(\subseteq)}(S)$
so the (canonical) isomorphism of algebras from
Claim~\ref{lem:calG(subseteq)(S)/calK(subseteq)S-canonical-isomorphism} takes the following form:
\begin{equation}\label{eqn:calG(subseteq)(S)-canonical-isomorphisms-new}
\tilde \calG_{(\subseteq)}(S) \cong \hat \calG_{(\subseteq)}(S),\quad x + \calK_{(\subseteq) S} \mapsto x + \calK,\quad x \in \calG_{(\subseteq)}(S).
\end{equation}

\begin{claim}\label{lem:projection-and-preimage-of-calK-ideals}
Let $S \subseteq \posints$ be finite.
For the linear map $\pi_S \colon \calG \to \calG_\subseteq(S)$
we have $\pi_S(\calK) = \calK_{{\subseteq}S}$.
\end{claim}
\begin{proof}
Because $(\pi_S)_{| \calG_\subseteq(S)} = \id_{|\calG_\subseteq(S)}$ and by Lemma~\ref{lem:calKsubseteqS}
$\calK_{{\subseteq}S} = \calG_\subseteq(S) \cap \calK$,
we infer that $\pi_S(\calK) \supseteq \calK_{{\subseteq}S}$.
For the reverse inclusion, let $x \in \calK$.
Fix any $y \in \calG_\subseteq(S)$.
Then by Claim~\ref{lem:projection}, $f(\pi_S(x) y) = f(x y) = 0$,
the last equality is true because $x \in \calK$.
Hence $\pi_S(x) \in \calK_{{\subseteq}S}$
so that $\pi_S(\calK) \subseteq \calK_{{\subseteq}S}$.
\end{proof}

For the linear map $\pi_S \colon \calG \to \calG_\subseteq(S) \subseteq \calG$
by Lemmas~\ref{lem:projection-and-preimage-of-calK-ideals} and~\ref{lem:calKsubseteqS},
$\pi_S(\calK) = \calK_{{\subseteq}S} = \calG_{\subseteq}(S) \cap \calK$,
so that we have the well-defined linear map (which we denote by $\hat \pi_S$)
\begin{equation}\label{eqn:quotient-projection}
\hat \pi_S \colon \hat \calG \to \hat \calG_\subseteq(S),\quad \hat \pi_S(x + \calK) = \pi_S(x) + \calK,\quad x \in \calG.
\end{equation}
 
It is easy to see that $(\hat \pi_S)_{| \hat \calG_\subseteq(S)} = \id_{|\hat \calG_\subseteq(S)}$.
In particular, $\hat \pi_S \colon \hat \calG \to \hat \calG_\subseteq(S)$ is surjective.
 
\begin{claim}\label{lem:quotion-decomposition-through-projeciton}
Let $S \subseteq \posints$ be finite. Then $\hat \calG = \hat \calG_\subseteq(S) \oplus (\hat \calG_\subseteq(S))^\perp$ via $x = \hat \pi_S(x) + (x - \hat \pi_S(x)),\, x \in \hat \calG$.
\end{claim}
\begin{proof}
%

First, let $x \in \hat \calG$. Write $x = y + \calK$ where $y \in \calG$.
Then $\hat \pi_S(x) = \pi_S(y) + \calK$ and $\pi_S(y) \in \calG_\subseteq(S)$.
We have $x - \hat \pi_S(x) = y - \pi_S(y) + \calK$.
By Claim~\ref{cor:projection-part-orthogonality}, $y - \pi_S(y) \in (\calG_\subseteq(S))^\perp$
so that $x - \hat \pi_S(x) \in (\hat \calG_\subseteq(S))^\perp$,
since the bilinear form on $\calG$ extends to $\hat \calG$ in~(\ref{eqn:f-on-hat-calG}).

So we only need to show that $\hat \calG_\subseteq(S) \cap (\hat \calG_\subseteq(S))^\perp = 0 (= \{\calK\})$.
Let $z$ belong to this intersection.
Write $z = t + \calK = t' + \calK$ where $t \in \calG_\subseteq(S)$
and $t' \in (\calG_\subseteq(S))^\perp$.
Then clearly $t - t' \in  \calK \subseteq (\calG_{\subseteq}(S))^\perp$,
and so $t = (t - t') + t' \in (\calG_\subseteq(S))^\perp$.
Thus $t \in \calG_\subseteq(S) \cap (\calG_\subseteq(S))^\perp = \calK_{{\subseteq}S} \subseteq \calK$,
the last inclusion holds by Claim~\ref{lem:calKsubseteqS}. 
Therefore $z = t + \calK = \calK$,
implying that $\hat \calG_\subseteq(S) \cap (\hat \calG_\subseteq(S))^\perp = 0$.
\end{proof}

Thus Claim~\ref{lem:quotion-decomposition-through-projeciton} allows us
to rightfully call $\hat \pi_S \colon \hat \calG \to \hat \calG_\subseteq(S)$
an orthogonal projection of $\hat \calG$ to $\hat \calG_\subseteq(S)$.

If $S, T \subseteq \posints$ are finite subsets,
then $\pi_S(\calPLG_{(\subseteq)}(T)) = \calPLG_{(\subseteq)}(S \cap T)$,
where the projection is surjective because
the restriction $(\pi_S)_{| \calPLG_{(\subseteq)}(S \cap T)} = \id_{|\calPLG_{(\subseteq)}(S \cap T)}$.
So by linearity we get $\pi_S(\calG_{(\subseteq)}(T)) = \calG_{(\subseteq)}(S \cap T)$.
Going to the quotients, we conclude that $\hat \pi_S(\hat \calG_{(\subseteq)}(T)) = \hat \calG_{(\subseteq)}(S \cap T)$.

\begin{claim}\label{lem:orthogonal-complement-equivalence-for-calG(subseteq)(S)}
Let $n \ge 2$ and $S \subseteq \posints$ be finite.
Then for any $x \in \calG_{(\subseteq)}(S)$, we have $x \in \calK_{(\subseteq) S}$
iff $f(x x_1 \cdots x_{n - 1}) = 0$ for all $x_1, \ldots, x_{n - 1} \in \calG_{(\subseteq)}(S)$.
\end{claim}
\begin{proof}

For $\Rightarrow$, it suffices to note that $\calG_{(\subseteq)}(S)$
is closed under multiplication (in $\calG$).
To prove $\Leftarrow$, note that $n - 2 \ge 0$ and for any $y \in \calG_{(\subseteq)}(S)$,
we have $f(x y) = f(x y U^{n - 2}) = 0$,
where $U = U_\emptyset$ in the $\calG_\subseteq(S)$ case and
$U = U_S$ in the $\calG(S)$ case,
so $x \in \calK_{(\subseteq) S}$.
\end{proof}

The primary goal of the various Claims above is to define the projection $\hat \pi_S \colon \calG \to \calG_\subseteq(S)$ to be used later and to prove Lemmas~\ref{lem:empty-annihilator-for-hat-calG(subseteq)(S)} and~\ref{lem:calK[k]-T(f,k,n)-xi(h)-criterion}.

\begin{lemma}\label{lem:empty-annihilator-for-hat-calG(subseteq)(S)}
Let $S \subseteq \posints$ be finite.
Then the annihilator of $\hat \calG_{(\subseteq)}(S)$ in $\hat \calG_{(\subseteq)}(S)$ is zero, i.e.,
if $x \in \hat \calG_{(\subseteq)}(S)$ and $f(x y) = 0$
for every $y \in \hat \calG_{(\subseteq)}(S)$, then $x$
is
the zero element of $\hat \calG_{(\subseteq)}(S)$, namely $\calK$.
\end{lemma}
\begin{proof}
Let $x \in \hat \calG_{(\subseteq)}(S)$ be
an element satisfying the hypothesis of the lemma.
Write $x = h_1 + \calK$ where $h_1 \in \calG_{(\subseteq)}(S)$.
By hypothesis, for every 
$y \in \hat \calG_{(\subseteq)}(S)$ we have $f(x y) = 0$.
Let $h_2 \in \calG_{(\subseteq)}(S)$
and put $y = h_2 + \calK \in \hat \calG_{(\subseteq)}(S)$.
Then $x y = h_1 h_2 + \calK$.
By the definition of $f$ on $\hat \calG$, $f(h_1 h_2) = f(x y) = 0$. 
Hence $h_1 \in \calK_{(\subseteq) S} \subseteq \calK$
where the last inclusion is true by Claims~\ref{lem:calKS} and~\ref{lem:calKsubseteqS}.
This implies that $x$ is the  zero element of $\hat \calG_{(\subseteq)}(S)$,
which is $\calK$.
\end{proof}

\begin{lemma}\label{lem:calK[k]-T(f,k,n)-xi(h)-criterion}
Let $k, r \ge 0$ and $n \ge \max(2, r)$.
Suppose the connection tensor $T(f, k, n)$ can be expressed as
\[
T(f, k, n) = \sum_{i = 1}^r a_i \bfx_i^{\otimes n},
\]
where $a_i \ne 0$ and $\bfx_i
\in \bbF^{\calPLG[k]}$ are nonzero and \emph{pairwise} linearly independent for $1 \le i \le r$.
Then for any $h \in \calG[k]$, we have $h \in \calK_{[k]}$ iff $\bfx_i(h) = 0, \, 1 \le i \le r$.
\end{lemma}
\begin{proof}
The lemma is clearly true for $r=0$. Let $r \ge 1$.
By Claim~\ref{lem:orthogonal-complement-equivalence-for-calG(subseteq)(S)},
$h \in \calK_{[k]}$ iff $f(h h_1 \cdots h_{n - 1}) = 0$ for all $h_1, \ldots, h_{n - 1} \in \calG[k]$.
In terms of $T(f, k, n)$, this is equivalent to $(T(f, k, n))(h, \cdot, \ldots, \cdot) = 0$
which is the same as
\[
\sum_{i = 1}^r a_i \, \bfx_i(h) \, \bfx_i^{\otimes (n - 1)} = 0.
\]
Now if $\bfx_i(h) = 0, \, 1 \le i \le r$, then this equality clearly holds.
Conversely, if this equality holds, then by Lemma~\ref{lem:tensor-powers-pairwise-implies-linear-independence-long},
$a_i \bfx_i(h) = 0$ but $a_i \ne 0$ so $\bfx_i(h) = 0, \,  1 \le i \le r$.
\end{proof}

We will also need the following lemma 
that classifies all subalgebras of $\bbF^m$ for $m \ge 0$.
A proof is given in
 subsection~\ref{subsec:bbF^m-subalgebras} of the Appendix. 
Recall that we allow zero algebras 
and require any subalgebra of an algebra to share the multiplicative identity.

\begin{lemma}\label{lem:F^m-subalgebras}
All subalgebras of $\bbF^m$, where $m \ge 0$, are of the following form:
For some partition $[m] = \bigsqcup_{i = 1}^s \calI_i$,
where $s \ge 0$, and $\calI_i \ne \emptyset$ for $i \in [s]$, 
the subalgebra has equal values on each $I_i$,
\[
\bbF^{(\calI_1, \ldots, \calI_s)} = \{ (c_1, \ldots, c_m) \in \bbF^m \mid \forall i \in [s], \ \forall j, j' \in \calI_i, \ c_j = c_{j'} \}.
\]
\end{lemma}

\subsection{Building an algebra isomorphism~\label{sec:build-alg-isom}}

In this part of the proof regarding $f$,
for an arbitrary fixed $k \ge 0$,
we assume that
there exists $n = n_k \ge 2$ such that $\symrank T(f, k, n) \le n - 1$.
We will pick an arbitrary such $n$ and call it $n_k$,
and then write $r_k = \symrank T(f, k, n_k)$.
(Note that this is weaker than the
uniform exponential boundedness in $k$ for $\symrank T(f, k, n)$
in Theorem~\ref{thm:uniform-exponential-rank-boundedness-converse},
nor do we require $f(K_0) = 1$ here.)

Then, for $n = n_k$, we can write 
\begin{equation}\label{eqn:T(f,k,n)-gen-dec}
T(f, k, n) = \sum_{i = 1}^{r_{k}} a_{k, n, i} \bfx_{k, n, i}^{\otimes n}.
\end{equation}
Then $a_{k, n, i} \ne 0$ and $0 \ne\bfx_{k, n, i} \in \bbF^{\calPLG[k]}$ 
are pairwise linearly independent for $1 \le i \le r_{k}$.

Define the linear map
\begin{equation}\label{eqn:Phi-map}
\Phi_{k, n} \colon \calG[k] \to \bbF^{r_{k}},\quad \Phi_{k, n}(h) = ( \bfx_{k, n, i}(h) )_{i= 1, \ldots, r_{k}}, \quad h \in \calG[k].
\end{equation}
We show that $\Phi_{k, n} \colon \calG[k] \to \bbF^{r_{k}}$
is a surjective algebra homomorphism,  
after a normalization step (to be carried out later).
Clearly, as $n \ge 2$, 
\[
h_1 \cdot h_2 \cdot h_3 \cdots h_n = (h_1 h_2) \cdot U_k \cdot h_3 \cdots h_n
\]
so
\[
f(h_1 \cdot h_2 \cdot h_3 \cdots h_n) = f((h_1 h_2) \cdot U_k \cdot h_3 \cdots h_n)
\]
for all $h_1, \ldots, h_n \in \calG[k]$.
(When $n = 2$ this is $f(h_1 h_2) = f((h_1 h_2) U_k)$.)
Therefore
\[
(T(f, k, n))(h_1, h_2, \cdot, \ldots, \cdot) =  (T(f, k, n))(h_1 h_2, U_k, \cdot, \ldots, \cdot)
\]
for all $h_1, h_2 \in \calG[k]$.
In terms of the decomposition in (\ref{eqn:T(f,k,n)-gen-dec}),
this is equivalent to 
\[
\sum_{i = 1}^{r_{k}} \, a_{k, n, i} \, \bfx_{k, n, i}(h_1) \, \bfx_{k, n, i}(h_2) \, \bfx_{k, n, i}^{\otimes (n - 2)} = \sum_{i = 1}^{r_{k}} a_{k, n, i} \, \bfx_{k, n, i}(h_1 h_2) \, \bfx_{k, n, i}(U_k) \, \bfx_{k, n, i}^{\otimes (n - 2)}.
\]
It follows that
\[
\sum_{i = 1}^{r_{k}} a_{k, n, i} \, \left( \bfx_{k, n, i}(h_1) \, \bfx_{k, n, i}(h_2) - \bfx_{k, n, i}(h_1 h_2) \, \bfx_{k, n, i}(U_k) \right) \, \bfx_{k, i}^{\otimes (n - 2)} = 0
\]
for any $h_1, h_2 \in \calG[k]$.
The condition $r_{k} \le n - 1$ allows
us to apply Lemma~\ref{lem:tensor-powers-pairwise-implies-linear-independence-long}.
Since $a_{k, n, i} \ne 0$ for $1 \le i \le r_{k}$,
we obtain that
\begin{equation}\label{eqn:char-prop-unsimp}
\bfx_{k, n, i}(h_1) \, \bfx_{k, n, i}(h_2) = \bfx_{k, n, i}(h_1 h_2) \, \bfx_{k, n, i}(U_k), \quad h_1, h_2 \in \calG[k], \quad 1 \le i \le r_{k}.
\end{equation}
Let $1 \le i \le r_{k}$.
Since $\bfx_{k, n, i} \ne 0$, there exists $h \in \calG[k]$ such that $\bfx_{k, n, i}(h) \ne 0$.
Substituting $h_1 = h_2 = h$ into (\ref{eqn:char-prop-unsimp}),
we infer that $\bfx_{k, n, i}(U_k) \ne 0, \, 1 \le i \le r_{k}$.

Therefore we can assume
in (\ref{eqn:T(f,k,n)-gen-dec})
 that each $\bfx_{k, n, i}$ is normalized so that $\bfx_{k, n, i}(U_k) = 1$ ($1 \le i \le r_{k}$).
Combined with this, condition (\ref{eqn:char-prop-unsimp}) becomes for $1 \le i \le r_{k}$,
\begin{equation}\label{eqn:char-prop-simp-full}
\begin{lcases}
\bfx_{k, n, i}(h_1 h_2) &= \bfx_{k, n, i}(h_1) \, \bfx_{k, n, i}(h_2), \quad h_1, h_2 \in \calG[k]; \\
\bfx_{k, n, i}(U_k) &= 1;
\end{lcases}
\end{equation}
so the linear functions $\bfx_{k, n, i} \colon \calG[k] \to \bbF, \, 1 \le i \le r_{k}$
are algebra homomorphisms. 
Then we have
\begin{gather*}
\Phi_{k, n}(g h) = ( \bfx_{k, n, 1}(g h), \ldots, \bfx_{k, n, r_k}(g h) ) = ( \bfx_{k, n, 1}(g) \, \bfx_{k, n, 1}(h), \ldots, \bfx_{k, n, r_k}(g) \, \bfx_{k, n, r_k}(h)) \\
= ( \bfx_{k, n, 1}(g), \ldots, \bfx_{k, n, r_k}(g) ) \cdot ( \bfx_{k, n, 1}(h), \ldots, \bfx_{k, n, r_k}(h) ) = \Phi_{k, n}(g) \Phi_{k, n}(h). 
\end{gather*}
So we have
\begin{gather*}
\Phi_{k, n}(g h) = \Phi_{k, n}(g) \Phi_{k, n}(h), \quad g, h \in \calG[k], \\
\Phi_{k, n}(U_k) = ( \bfx_{k, n, 1}(U_k) \ldots, \bfx_{k, n, r_k}(U_k) ) = \underbrace{(1, \ldots, 1)}_{r_{k} \text{ times}} \in \bbF^{r_{k}},
\end{gather*}
and therefore $\Phi_{k, n} \colon \calG[k] \to \bbF^{r_{k}}$ is an algebra homomorphism.
We now prove its surjectivity.
Clearly, $\im(\Phi_{k, n})$ is a subalgebra of $\bbF^{r_{k}}$.
By Lemma~\ref{lem:F^m-subalgebras},
we may assume that $\im(\Phi_{k, n})$
has the form $\bbF^{(\calI_1, \ldots, \calI_s)}$ for some partition
$\{\calI_1, \ldots, \calI_s\}$ of $[r_{k}]$.
The pairwise linear independence of $\bfx_{k, n, i}$
for $1 \le i \le r_{k}$ implies that for any $1 \le i_1 < i_2 \le r_{k}$,
we have $\bfx_{k, n, i_1} \ne \bfx_{k, n, i_2}$,
so there exists $h \in \calG[k]$ such that $\bfx_{k, n, i_1}(h) \ne \bfx_{k, n, i_2}(h)$.
Since each $\calI_i \ne \emptyset$,
it follows that $|\calI_i| = 1$ for $1 \le i \le s$.
Hence $\im(\Phi_{k, n}) = \bbF^{(\{ 1 \}, \ldots, \{ r_k \})} = \bbF^{r_{k}}$.
We have shown that $\Phi_{k, n} \colon \calG[k] \to \bbF^{r_{k}}$ is surjective.

This results in the following lemma.
\begin{lemma}\label{lem:Phi-surjective-homomorphism}
Let $k \ge 0$. The constructed map $\Phi_{k, n} \colon \calG[k] \to \bbF^{r_{k}}$ defined in~(\ref{eqn:Phi-map}) is a surjective algebra homomorphism,
after the  normalization to set $\bfx_{k, n, i}(U_k) = 1$ for $1 \le i \le r_k$.
%
\end{lemma}

Next, by $r_{k} \le n - 1$ and $n \ge 2$, 
clearly $n \ge \max(2, r_{k})$, 
so Lemma~\ref{lem:calK[k]-T(f,k,n)-xi(h)-criterion} applies. So we have
\[
\ker \Phi_{k, n} = \{ h \in \calG[k] \mid \bfx_{k, n, i}(h) = 0,\,
1 \le i \le r_{k}  \} = \calK_{[k]},
\]
where the first equality is by the definition of $\Phi_{k, n}$,
and the second equality is by Lemma~\ref{lem:calK[k]-T(f,k,n)-xi(h)-criterion}.
Note that by Claim~\ref{lem:calKS}, we have $\calK_{[k]} =
\calG[k] \cap \calK$.
%
Then $\Phi_{k, n} \colon \calG[k] \to \bbF^{r_k}$
factors through $\calG[k] / \ker \Phi_{k, n} = \calG[k] / \calK_{[k]} = \tilde \calG[k]$,
inducing an algebra isomorphism
\[
\tilde \Phi_{k, n} \colon \tilde \calG[k] \to \bbF^{r_{k}}, \quad \tilde \Phi_{k, n}(h + \calK_{[k]}) = ( \bfx_{k, n, 1}(h), \ldots, \bfx_{k, n, r_k}(h) ),\quad h \in \calG[k].
\]
It follows that $\dim \tilde \calG[k] = \dim \bbF^{r_{k}} = r_{k}$.
In particular, $\tilde \calG[k]$ is a finite dimensional algebra.
Applying Lemma~\ref{lem:quotient-space-dimension-long},
we get $\dim \tilde \calG[k] = \dim \Span \{ \bfx_{k, n, i} \}_{i = 1}^{r_{k}}$.
Then $\dim \Span \{ \bfx_{k, n, i} \}_{i = 1}^{r_{k}} = r_{k}$
implying that $\bfx_{k, n, i}$,  $1 \le i \le r_{k}$, are linearly independent.
(Note that we started off only assuming they are nonzero and pairwise linearly independent.)
We formalize some of the results obtained above.
\begin{lemma}\label{lem:Phi(k)-tilde-calG[k]-isomorphism}
Let $k \ge 0$. Assume there exists $n = n_k \ge 2$
such that $r_{k} = \symrank T(f, k, n_k) \le n_k - 1$.
Then the constructed map
\[
\tilde \Phi_{k, n} \colon \tilde \calG[k] \to \bbF^{r_{k}}, \quad \tilde \Phi_k(h + \calK_{[k]}) = ( \bfx_{k, n, 1}(h), \ldots, \bfx_{k, n, r_k}(h) ),\quad h \in \tilde \calG[k].
\]
is an algebra isomorphism and $\dim \tilde \calG[k] = r_{k}$.
\end{lemma}

Composing $\tilde \Phi_k \colon \tilde \calG[k] \to \bbF^{r_{k}}$
with the canonical algebra isomorphism between $\tilde \calG[k]$ and $\hat \calG[k]$
given in (\ref{eqn:calG(subseteq)(S)-canonical-isomorphisms-new}),
we have an algebra isomorphism $\hat \Phi_k \colon \hat \calG[k] \to \bbF^{r_{k}}$.
In particular, $\dim \hat \calG[k] = \dim \tilde \calG[k] = r_{k}$.
\begin{corollary}\label{cor:Phi(k)-hat-calG[k]-isomorphism}
With the same assumption as in Lemma~\ref{lem:Phi(k)-tilde-calG[k]-isomorphism},
the  map
\[
\hat \Phi_{k, n} \colon \hat \calG[k] \to \bbF^{r_{k}}, \quad \hat \Phi_{k, n}(h + \calK) = ( \bfx_{k, n, 1}(h), \ldots, \bfx_{k, n, r_k}(h) ),\quad h \in \hat \calG[k].
\]
is an algebra isomorphism and  $\dim \hat \calG[k] = r_{k}$.
\end{corollary}

Note that if $S \subseteq \posints$ is finite and $|S| = k$,
there are natural isomorphisms between $\tilde \calG(S)$ and $\tilde \calG[k]$
and also between $\hat \calG(S)$ and $\hat \calG[k]$,
both resulting from any bijective map between $S$ and $[k]$.
As a result, we conclude the following.
\begin{corollary}\label{cor:dimension-of-tilde-hat-calG(S)}
%
Let $k \ge 0$ and $S \subseteq \posints$ with $|S| = k$.
Suppose there exists some $n = n_k \ge 2$ so that $\symrank T(f, k, n) \le n - 1$. 
Let $r_k = \symrank T(f, k, n)$.
Then $\tilde \calG(S) \cong \hat \calG(S) \cong \bbF^{r_{k}}$
and $\dim \tilde \calG(S) = \dim \hat \calG(S) = r_{k}$.
In particular, the value $r_k$ is independent of the choice of $n$.
\end{corollary}

\subsection{One \texorpdfstring{$n$}{n} implies for all \texorpdfstring{$n$}{n}~\label{sec:one-implies-all}}

Let $n_k$ retain the same meaning
as in Lemma~\ref{lem:Phi(k)-tilde-calG[k]-isomorphism},
and let $r = r_k =  \symrank(T(f, k, n_k))$.
For any $h \in \calG[k]$, clearly
$h = h U_k^{n_k - 1}$ so $f(h) = f(h U_k^{n_k - 1})$.
As $\bfx_{k, n_k, i}(U_k) = 1, \, 1 \le i \le r$,
\[
\begin{split}
f(h) &= f(h U_k^{n_k - 1}) = (T(f, k, n_k))(h, U_k, \ldots, U_k) = \sum_{i = 1}^r a_{k, n_k, i} (\bfx_{k, n_k, i}(U_k))^{n_k - 1} \bfx_{k, n_k, i}(h) \\
&= \sum_{i = 1}^r a_{k, n_k, i} \bfx_{k, n_k, i}(h)
\end{split}
\]
for any $h \in \calG[k]$.
Hence $f_{|\calG[k]} = \sum_{i = 1}^r a_{k, n_k, i} \bfx_{k, n_k, i}$, i.e.,
$f_{|\calG[k]}$ is a linear combination of $r$ algebra 
homomorphisms $\bfx_{k, n_k, i} \colon \calG[k] \to \bbF$, for $1 \le i \le r$.
In particular, applying to the product $h_1 \cdots h_n$,
for any $n \ge 0$ and  any  $h_1, \ldots, h_n \in \calG[k]$,
(note that this $n$ is arbitrary, not only for  those $n$ satisfying
the requirements for the choice of $n_k$)
\[
f(h_1 \cdots h_n) = \sum_{i = 1}^r a_{k, n_k, i} \bfx_{k, n_k, i}(h_1 \cdots h_n) = \sum_{i = 1}^r a_{k, n_k, i} \bfx_{k, n_k, i}(h_1) \cdots  \bfx_{k, n_k, i}(h_n).
\]
(When $n = 0$, we view it as $f(U_k) = \sum_{i = 1}^r a_{k, n_k, i} \bfx_{k, n_k, i}(U_k) = \sum_{i = 1}^r a_{k, n_k, i}$.)
Hence
\begin{equation}\label{eqn:T(f,k,n)-char-dec}
T(f, k, n) = \sum_{i = 1}^r a_{k, n_k, i} \bfx_{k, n_k, i}^{\otimes n}
\end{equation}
for all $n \ge 0$.
(When $n = 0$, 
(\ref{eqn:T(f,k,n)-char-dec}) is still valid as
$T(f, k, 0) = f(U_k) = \sum_{i = 1}^r a_{k, n_k, i} = \sum_{i = 1}^r a_{k, n_k, i} \bfx_{k, n_k, i}^{\otimes 0}$
where the last equality is true as $\bfx_{k, n_k, i}^{\otimes 0} = 1$.)

As shown before, $\bfx_{k, n_k, i}$ where $1 \le i \le r$ are linearly independent.
Then by Lemma~\ref{lem:tensor-powers-linear-independence-symmetric-rank-and-uniqueness-long} 
applied to (\ref{eqn:T(f,k,n)-char-dec}),
we get $\symrank(T(f, k, n)) = r$ for all $n \ge 2$;
and the decomposition (\ref{eqn:T(f,k,n)-char-dec})
is actually unique up to a permutation for $n \ge 3$.

To summarize, this leads to the following. 

\begin{theorem}\label{thm:T(f,k,n)-rk-unif-dec-uniq}
Let $k \ge 0$.
Assume that for some $n = n_k \ge 2$, $\symrank T(f, k, n) = r \le n - 1$.
Then the following hold:
\begin{enumerate}
\item $\symrank T(f, k, n) = r$ for every $n \ge 2$.
\item There exist $r$ linearly independent algebra homomorphisms $\bfx_i \colon \calG[k] \to \bbF$,
and $a_1, \ldots, a_r \in \bbF \setminus \{ 0 \}$ such that $f_{|\calG[k]} = \sum_{i = 1}^r a_i \bfx_i$;
also for every $n \ge 0$,
\begin{equation}\label{eqn:T(f,k,n)-unif-dec}
T(f, k, n) = \sum_{i = 1}^r a_i \bfx_i^{\otimes n}.
\end{equation}
Moreover, for any $n \ge 3$, any expression of $T(f, k, n)$ as $\sum_{i = 1}^r b_i \bfy_i^{\otimes n}$,
where $\bfy_i \colon \calG[k] \to \bbF$ are linear maps,
is a permutation of the sum in (\ref{eqn:T(f,k,n)-unif-dec}).
\end{enumerate}
\end{theorem}

We remark that this  is  a nontrivial statement:
The existence of some $n_k$ has produced
a uniform expression for the tensors $T(f, k, n)$
all the way down to $n = 0$.

\subsection{Putting things together}

From now on, we assume that for every $k \ge 0$,
there exist some $n = n_k \ge 2$ such that $\symrank T(f, k, n) \le n - 1$.
For every $k \ge 0$, we pick an arbitrary such $n$, call it $n_k$,
and let $r_k = \symrank T(f, k, n_k)$.

Having developed the theory in a more general setting,
we can now follow the proof in~\cite{Freedman-Lovasz-Schrijver-2007} closely.
As mentioned before, developing this theory in a more general setting
is necessary because $f(K_1) = 0$  is possible which makes
 the normalization step from~\cite{Freedman-Lovasz-Schrijver-2007} infeasible. Now the main difference from~\cite{Freedman-Lovasz-Schrijver-2007} starting from this point is that many of our derivations will additionally contain units of the form $u_S$ for various finite $S \subseteq \posints$ because we cannot ensure that $u_S = u_\emptyset$. 
We will be interested in the idempotent elements of $\hat \calG$.
For two elements  $p, q \in \hat \calG$,
we say that $q$ \textit{resolves} $p$, if $p q = q$.
We also say equivalently $p$ is resolved by $q$. 
It is clear that the binary relation
\textit{resolves} is antisymmetric and transitive and,
when restricted to idempotents, reflexive.
Furthermore, it is easy to see that the binary 
relation \textit{resolves} on $\hat \calG$ has the following properties:

\begin{enumerate}
\item The idempotent $0 = \calK$ resolves everything and $1 = u_\emptyset = 
U_\emptyset + \calK = K_0 + \calK$ is resolved by everything.
\item If $a b = 0$ and $c$ resolves both $a$ and $b$, then $c = 0$;
\item If $a$ resolves $b$, then $c$ resolves  $a$ iff $c$ resolves $a b$.
\end{enumerate}

In the algebra $\bbF^r$ ($r \ge 0$), 
the idempotents are 0-1 tuples in $\bbF^r$, and for idempotents
$q = (q_1, \ldots, q_r)$ and $p = (p_1, \ldots, p_r)$,
$q$ resolves $p$ iff $q_i=1$ implies $p_i=1$.

Let $S$ be a finite subset of $\posints$ with $|S|=k$, and set $r = r_{k}$
as above.
By Corollary~\ref{cor:dimension-of-tilde-hat-calG(S)},
$\hat \calG(S) \cong \bbF^r$ as algebras,
so $\hat \calG(S)$ has a (uniquely determined idempotent) basis $\calP_S = \{ p_1^S, \ldots, p_r^S \}$
such that $(p_i^S)^2 = p_i^S$ and $p_i^S p_j^S = 0$ for $i \ne j$.
These correspond to the canonical basis $\{e_i\}_{1 \le i \le r}$ of $\bbF^r$
under this isomorphism.
For $i \ne j$, we have $\langle p_i^S, p_j^S \rangle = f(p_i^S p_j^S) = 0$.
Furthermore, for all $1 \le i \le r$,
\begin{equation}\label{eq:nonzero-idempotent}
f(p_i^S) = f((p_i^S)^2) = \langle p_i^S, p_i^S \rangle \ne 0,
\end{equation}
otherwise $\hat \calG(S)$ contains a nonzero element orthogonal to $\hat \calG(S)$
with respect to the bilinear form $\langle \cdot, \cdot \rangle$
restricted to $\hat \calG(S) \times \hat \calG(S)$, 
contradicting Lemma~\ref{lem:empty-annihilator-for-hat-calG(subseteq)(S)}.
We will call the elements $p_i^S \in \calP_S$ basic idempotents. 

%

We denote by $\calP_{T, p}$ the set of all idempotents in $\calP_T$
that resolve a given element $p \in \hat \calG$.
If $p \in \calP_S$ and $S \subset T$ and $|T| = |S| + 1$,
then the number of elements in $\calP_{T, p}$
will be called the degree of $p \in \calP_S$, and denoted by $\deg(p)$.
Obviously, this value is independent of
which $(|S| + 1)$-element superset $T$ of $S$ we take.

For any $q \in \hat \calG(T)$, we have $q u_{T \setminus S} = q$. (Here by definition, $u_{T \setminus S} = U_{T \setminus S} + \calK \in \hat \calG(T \setminus S) \subseteq \hat \calG(T) \subseteq \hat \calG$.) 
It follows that for any $S \subseteq T$ and $p \in \hat \calG$,
we have  $q$ resolves $p$ iff $q$ resolves $p u_{T \setminus S}$,
since $q p = q u_{T \setminus S} p = q p u_{T \setminus S}$.
It is also important to point out that
an element in $\hat \calG(S)$ is an idempotent in $\hat \calG(S)$
iff it is an idempotent in $\hat \calG$.

%
%

\begin{claim}\label{clm:idempotent-is-sum-of-its-resolving-basic-idempotents}
Let $x$ be any idempotent element of $\hat \calG(S)$.
Then $x$ is the sum of exactly those idempotents in $\calP_S$ that resolve it,
\[
x = \sum_{p \in \calP_{S, x}} p.
\]
\end{claim}
\begin{proof}
Let $k = |S|$, and $r = r_{k}$.  By the isomorphism  $\hat \calG(S) \cong \bbF^{r}$, 
every 0-1 tuple $x = (x_1, \ldots, x_r) \in \bbF^r$
is the sum $\sum_{i=1}^r  x_i e_i$.
\end{proof}

In particular
\begin{equation}\label{eqn:uS-decomposition-in-idempotent-basis}
u_S = \sum_{p \in \calP_S} p,
\end{equation}
since  $u_S \in \hat \calG(S)$ corresponds
to the all-1 tuple in $\bbF^r$.

\begin{claim}\label{clm:subset-inclusion-and-uniqueness-of-resolving-basic-idempotents}
Let $S \subseteq T$ be two finite sets.
Then every $q \in \calP_T$ resolves exactly one element of $\calP_S$.
\end{claim}
\begin{proof}
Let $k' = |T|$ and $r' = r_{k'}$.
Consider the idempotents $pu_{T \setminus S}$
(which could be 0) under the isomorphism $\hat \calG(T) \cong \bbF^{r'}$,
where $p \in \calP_S$.
We can write the 0-1 tuple corresponding to $pu_{T \setminus S}$
in $\bbF^{r'}$ as the sum of those canonical basis $0$-$1$ vectors.
Recall that for any $q \in \calP_T$,
$q$ resolves $p$
iff $q$  resolves $pu_{T \setminus S}$.
Note that $pu_{T \setminus S}$ must have disjoint positions with entry 1 
for distinct $p \in \calP_S$, and the sum $\sum_{p \in \calP_S} 
pu_{T \setminus S} = u_S u_{T \setminus S} = u_T$ is the all-1 
tuple in $\bbF^{r'}$. 
Thus each $q \in \calP_T$ resolves exactly one $p \in \calP_S$.
\end{proof}

\begin{claim}\label{clm:projection-removal-inside-f}
Let $T$ and $U$ be finite sets, and let $S = T \cap U$.
If $x \in \hat \calG(T)$ and $y \in \hat \calG(U)$, then
\[
f(x y) = f(\hat \pi_S(x) y).
\]
\end{claim}
\begin{proof}
%
For every $T$-labeled graph $G_1$ and $U$-labeled graph $G_2$,
the graphs $G_1 G_2$ and $\pi_S(G_1) G_2$ are isomorphic as unlabeled graphs.
Hence $f(G_1 G_2) = f(\pi_S(G_1) G_2)$ as $f$ ignores labels.
Then we extend the equality from $\calPLG$ by linearity to $\calG$
and after that proceed to the quotient $\hat \calG = \calG / \calK$
using the definition of $f$ on $\hat \calG$.
\end{proof}

We remarked in~(\ref{eq:nonzero-idempotent})  that $f(p) \ne 0$ for any $p \in \calP_S$.
\begin{claim}\label{clm:projection-formula}
Let $S \subseteq T$ be two finite sets.
If $q \in \calP_T$ resolves $p \in \calP_S$, then
\[
\hat \pi_S(q) = \frac{f(q)}{f(p)} p.
\]
\end{claim}
\begin{proof}
Note that $q \in \hat \calG(T)$.
Since $S \subseteq T$, it follows that $\hat \pi_S(q) \in \hat \calG(S)$.
Because the only element from $\hat \calG(S)$ orthogonal to $\hat \calG(S)$
with respect to the dot product $\langle \cdot, \cdot \rangle$
restricted to $\hat \calG(S) \times \hat \calG(S)$ is $0$ (by Lemma~\ref{lem:empty-annihilator-for-hat-calG(subseteq)(S)}),
it suffices to show that both sides give
the same dot product with every basis element in $\cal P_S$.
For any $p' \in \calP_S \setminus \{ p \}$, we have $p' p = 0$ so $p' q = p' p q = 0$.
By Claim~\ref{clm:projection-removal-inside-f},
this implies that
\[
\langle p', \hat \pi_S(q) \rangle = f(p' \hat \pi_S(q)) = f(p' q) = 0 = \langle p', \frac{f(q)}{f(p)} p \rangle.
\]
Furthermore,
\[
\langle p, \hat \pi_S(q) \rangle = f(p \hat \pi_S(q)) = f(p q) = f(q) = \langle p, \frac{f(q)}{f(p)} p \rangle.
\]
This proves the claim.
\end{proof}

\begin{claim}\label{clm:swaps-within-f}
Let $T$ and $U$ be finite sets and let $S = T \cap U$.
Then for any $p \in \calP_S, q \in \calP_{T, p}$ and $r \in \hat \calG(U)$ we have
\[f(q r) = \frac{f(q)}{f(p)} f(rp).\]
\end{claim}
\begin{proof}
By 
Claims~\ref{clm:projection-removal-inside-f}~and~\ref{clm:projection-formula},
\[
f(q r) = f(\hat \pi_S(q) r) = \frac{f(q)}{f(p)} f(rp).
\]
\end{proof}

\begin{claim}\label{clm:basic-idempotents-nonzero-product}
Let $T$ and $U$ be finite sets and let $S = T \cap U$.
If both $q \in \calP_T, r \in \calP_U$ resolve $p \in \calP_S$, then $q r \ne 0$.
\end{claim}
\begin{proof}
By Claim~\ref{clm:swaps-within-f},
\[
f(q r) = \frac{f(q)}{f(p)} f(r p) = \frac{f(q)}{f(p)} f(r) \ne 0.
\]
\end{proof}

\begin{claim}\label{clm:degree-monotonicity}
If $S \subseteq T$, and $q \in \calP_T$ resolves $p \in \calP_S$, then $\deg(q) \ge \deg(p)$.
\end{claim}
\begin{proof}
It suffices to show this in the case when $|T| = |S| + 1$.
Let $U \subset \posints$ be any $(|S| + 1)$-element superset of $S$ different from $T$;
suppose this is the disjoint union $U = S \sqcup \{ a \},\, a \notin T$. 
Let $Y$ be the set of elements in $\calP_U$ resolving $p u_{\{ a \}}$
(equivalently, resolving $p$, because
every $r \in \calP_U$ resolves $u_{\{ a \}}$ as $a \in U$).
Then $p u_{\{ a \}} = \sum_{r \in Y} r$
by Claim~\ref{clm:idempotent-is-sum-of-its-resolving-basic-idempotents}.
Here $|Y| = \deg(p)$.
Furthermore, we have
\begin{equation}\label{eqn:qr-summing-to-qu_(a)}
\sum_{r \in Y} q r = q \sum_{r \in Y} r = q p u_{\{ a \}} = q u_{\{ a \}}.
\end{equation}
Each term $q r$ on the left hand side is nonzero
by Claim~\ref{clm:basic-idempotents-nonzero-product},
and since the terms are all idempotent,
each of them is a sum of one or more elements of $\calP_{T \cup U}$.
Furthermore, if $r, r' \in Y$ ($r \ne r'$),
then we have the orthogonality relation
\[
(q r) (q r') = q (r r')  = 0,
\]
so the sets of basic idempotents of $\calP_{T \cup U}$
in the expansion of each term are pairwise disjoint.
Therefore the expansion $\sum_{r \in Y} q r$
in $\calP_{T \cup U}$ contains at least $|Y| = \deg(p)$ terms.
On the right hand side of~(\ref{eqn:qr-summing-to-qu_(a)}), for any $z \in \calP_{T \cup U}$,
$z$ resolves $q$ iff $z$ resolves $q u_{\{ a \}}$ since $a \in U$.
Thus the number of terms in the expansion of $q u_{\{ a \}}$
in the basis $\calP_{T \cup U}$ is precisely $\deg(q)$ by definition.
Thus, $\deg(q) \ge |Y| = \deg(p)$.
The claim is proved.
\end{proof}

\subsection{Bounding the expansion}\label{subsec:Bounding-expansion}

At this point, we finally assume that all the conditions of
Theorem~\ref{thm:uniform-exponential-rank-boundedness-converse-strengthened}
are satisfied, i.e.,
$f(K_0) = 1$ and there is an integer $q \ge 0$ such that for 
every $k \ge 0$ there exists $n = n_k \ge 2$ satisfying
$r_{k, n} = \symrank T(f, k, n) \le \min(n - 1, q^k)$.
In particular $r_{0, n} \le q^0 = 1$ for $n = n_0 \ge 2$.
Clearly, $r_{0, n} \ne 0$ since $f(K_0) = 1$ so $r_{0, n} = 1$.
By Proposition~\ref{prop:graph-paramater-multiplicativity-criterion}, $f$ is multiplicative.

Next, from $f(K_0) = 1$,
we have $U_\emptyset = K_0 \notin \calK$
so that $u_\emptyset = U_\emptyset + \calK$ is a nonzero identity
in $\hat \calG(\emptyset) = \hat \calG[0] \ne 0$.
As $u_\emptyset$ is the sum of all basic idempotents in $\calG(\emptyset)$
we infer that $\calP_\emptyset \ne 0$.
Hence there is at least one basic idempotent.

 
If for any finite $S \subseteq \posints$,
a basic idempotent $p \in \calP_S$ has degree $D \ge 0$,
then for any superset $T \subseteq \posints$ of $S$ with $|T| = |S| + 1$, 
there are $D$ basic idempotents resolving $p$.
Let $S \subseteq T$, $t = |T \setminus S|$,
and $T \setminus S = \{u_1, u_2, \ldots, u_t\}$.
For each $1 \le i \le t$, we can pick $D$ basic idempotents
$q^{u_i}_j \in \calP_{S \cup \{u_i\}}$ resolving $p$, where $1 \le j \le D$.
For any mapping $\phi \colon \{1, \ldots, t\} \rightarrow \{1, \ldots, D\}$,
we can form the product $q_\phi = \prod_{i = 1}^t  q^{u_i}_{\phi(i)}$.
If $t = 0$, we assume $q_\phi = p$.
These are clearly idempotents resolving $p$.
If $\phi \ne \phi'$ then for some $i$, 
we have the orthogonality relation $q^{u_i}_{\phi(i)} q^{u_i}_{\phi'(i)} = 0$.
Thus $q_\phi q_{\phi'} = 0$. 
Also by applying Claim~\ref{clm:swaps-within-f} $t$ times,
\begin{equation}\label{f(q-phi)-factored-into-product-and-nonzero-1stpass}
f(q_\phi) = f(q_\phi p) = f(\prod_{i = 1}^t q^{u_i}_{\phi(i)} p ) = (\prod_{i = 1}^t \frac{f(q^{u_i}_{\phi(v)})}{f(p)}) f(p) \ne 0,
\end{equation}
and so $q_\phi \ne 0$.
Thus the set $\{q_\phi \mid \phi \colon \{1, \ldots, t\} \rightarrow \{1, \ldots, D\} \}$ is linearly independent. 
This implies that the dimension of $\hat \calG(T)$ over $\bbF$
is at least $D^{t} = D^{|T| - |S|}$.
But by Corollary~\ref{cor:dimension-of-tilde-hat-calG(S)}
and the hypothesis of Theorem~\ref{thm:uniform-exponential-rank-boundedness-converse-strengthened}
we also have the upper bound $q^{|T|}$.
If $D > q$, this leads to a contradiction if $|T|$ is large.
It follows that $D \le q$,
i.e., the degrees of basic idempotents for any $S$
and any $p \in \calP_S$ are bounded by $q$.
Let $D \ge 0$ denote the maximum degree
over all such $S$ and $p \in \calP_S$,
and suppose it is attained at some particular $S$ and $p \in \calP_S$.
We now fix this $S$ and $p$.
Note that for the existence of $D$
we also use the existence of a basic idempotent.

For $u \in \posints \setminus S$, let $q_1^u, \ldots, q_D^u$
denote the elements of $\calP_{S \cup \{ u \}}$ resolving $p$.
Note that for $u, v \in \posints \setminus S$,
there is a natural isomorphism between $\hat \calG(S \cup \{ u \})$ and $\hat \calG(S \cup \{ v \})$
(induced by the map that fixes $S$ pointwise and maps $u$ to $v$),
and we may choose the indexing so that
$q_i^u$ corresponds to $q_i^v$ under this isomorphism.


Now for any finite set $T \supseteq S$
all basic idempotents in $\calP_T$ that resolve $p$ can be described.
To describe these,
 let $V = T \setminus S$, and for every map $\phi \colon V \to \{ 1, \ldots, D \}$, we 
define as before
\begin{equation}\label{eqn:q-phi-factored-into-product}
q_\phi = \prod_{v \in V} q_{\phi(v)}^v.
\end{equation}

We have shown that these are linearly independent.

\begin{claim}\label{clm:calP_(T,p)-resolved-by-q_phi}
\[
\calP_{T, p} = \{ q_\phi \colon \phi \in \{ 1, \ldots, D \}^V \}.
\]
\end{claim}
\begin{proof}
We prove this by induction on the cardinality of $V = T \setminus S$.
For $|V| = 0, 1$ the assertion is trivial.
Suppose that $|V| > 1$.
Pick any $u \in V$,  let $U = S \cup \{ u \}$
and $W = T \setminus \{ u \}$; thus $U \cap W = S$.
By the induction hypothesis, the basic idempotents in $\calP_W$ resolving $p$
are elements of the form $q_\psi$,
for $\psi \in \{1, \ldots, D \}^{W}$.

Let $r$ be one of these.
By Claim~\ref{clm:basic-idempotents-nonzero-product},
$r q_i^u \ne 0$ for any $1 \le i \le D$, and clearly resolves $r$.
We can write $r q_i^u$ as a sum of basic idempotents in $\calP_T$
resolving it,
and it is easy to see that these also resolve $r$ (as 
\emph{resolve} is transitive).
For each  $r q_i^u$ the sum is nonempty as $r q_i^u \ne 0$.
Furthermore, the sets of basic idempotents occurring
in the expressions for $r q_i^u$ and $r q_j^u$ ($i \ne j$) are disjoint;
this follows from item 2 stated at the beginning of this subsection,
and $q_i^u q_j^u =0$.
If the sum  for any $r q_i^u$ has more than one basic idempotent,
then $r$ would have degree $>D$, violating the maximality of $D$.
So each $r q_i^u$ must be a basic idempotent in $\calP_T$ itself.

Each $r \in \calP_{W}$ resolves $p$ iff $r$ resolves $p u_{W}$.
Hence $p u_{W} = \sum_{r \in \calP_{W, p}} r$.
Also $p u_{\{ u \}} = \sum_{i = 1}^D q_i^u$.
Therefore we have
\[
p u_{T} = p u_{W} p u_{\{ u \}} = \sum_{r \in \calP_{W, p}, 1 \le i \le D} r q_i^u
\]
i.e., the basic idempotents $r q_i^u$ ($r \in \calP_{W, p},\, 1 \le i \le D)$
form the set of basic idempotents in $\calP_{T}$ resolving $p u_{T}$,
which is equivalent to resolving $p$.
It follows that these are all the elements of $\calP_{T, p}$.
This proves the claim.
\end{proof}

It is immediate from the definition that
an idempotent $q_\phi$ resolves $q_i^v u_V$ (equivalently $q_i^v$) iff $\phi(v) = i$.
Hence it also follows that
\[
q_i^v u_V= \sum_{\phi \colon \phi(v) = i} q_\phi.
\]
By the same reason, it also follows that for $u, v \in V$, $u \ne v$,
and any $1 \le i, j \le D$, 
\begin{equation}\label{eqn:q-i-v-times-u-v-factored-into-product-pairs}
q_i^u q_j^v u_V= \sum_{\substack{\phi \colon \phi(u) = i \\ 
\quad \phi(v) = j}} q_\phi.
\end{equation}

\subsection{Constructing the target graph}

Now we can define $H$ as follows.
Let $S$ and $p$ be fixed as above. 
For any $u \in \posints \setminus S$, let $\{q_1^u, \ldots, q_D^u\}$ 
be defined as in subsection~\ref{subsec:Bounding-expansion}.

Let $H$ be the looped complete graph on $V(H) = \{ 1, \ldots, D \}$.
We have to define the node weights and edge weights.
For every $i \in V(H)$, let
\[
\alpha_i = \frac{f(q_i^u)}{f(p)}
\]
be the weight of node $i$~\footnote{For $\bbF = \bbR$,
if we require the positive semidefiniteness
of the connection matrices $M(f, k)$ for $k \ge 0$,
then since $p$ and $q_i^u$ are basic idempotents,
$f(p) = f(p^2) > 0$ and similarly $f(q_i^u) > 0$.
Thus $\alpha_i > 0$, and so we recover the positive vertex weight case; 
see~\cite{Freedman-Lovasz-Schrijver-2007}.}. 
This definition does not depend on the choice of $u$,
because if $v \in \posints \setminus S$
and $v \ne u$, then the isomorphism
from $\hat \calG(S \cup \{ u \})$ to $\hat \calG(S \cup \{ v \})$ 
(induced by the map that fixes $S$ and maps $u$ to $v$),
will send $q_i^u$ to $q_i^v$.

%
%

Let $u, v \in \posints \setminus S,\, u \ne v$, and let $W = S \cup \{ u, v \}$.
Let $K_{u v}$ denote the graph on the vertices $u$ and $v$
that are correspondingly labeled $u$ and $v$,
and has only one edge connecting $u$ and $v$.
Let $k_{u v} = K_{u v} + \calK$ denote the corresponding element of $\hat \calG(\{ u, v \})$.
We can express $p k_{u v}$ as a linear combination
of the basic idempotents from $\calP_W$.
Note that $r \in \calP_{W}$ resolves $p$ iff $r$ resolves $p u_{\{ u,v \}}$,
thus $p u_{\{ u, v \}} = \sum_{r \in \calP_{W, p}} r$.
So if $r' \in \calP_W \setminus \calP_{W, p}$, we have $r' p = r' u_{\{ u, v \}} p =0$.
Thus $r' p k_{u v} = 0$. It follows that $p k_{u v}$ is a linear 
combination of the basic idempotents from the subset $\calP_{W, p}$.
We write this unique expression
\[
p k_{u v} = \sum_{i, j = 1}^D \beta_{i j} q_i^u q_j^v.
\]
This defines (by the uniqueness) the weight $\beta_{i j}$ of the edge $i j$.
Note that $\beta_{i j} = \beta_{j i}$ for all $i, j$, since $p k_{u v} = p k_{v u}$.

We prove that this weighted graph $H$ gives the desired homomorphism function.

\begin{claim}\label{clm:f-equals-hom(,H)}
For every finite graph $G$, $f(G) = \hom(G, H)$.
\end{claim}
\begin{proof}
Let $V$ be a finite subset of $\posints$ disjoint from $S$ of cardinality $|V(G)|$.
We label $V(G)$ by $V$ thus making $G$ a $V$-labeled graph, so now $G \in \calG(V)$.
Since $f$ ignores labels, we may identify $V$ and $V(G)$, and assume $V(G) = V$.
Now we take $T = S \sqcup V$,
thus $V = T \setminus S$. This defines $q_\phi$
as in subsection~\ref{subsec:Bounding-expansion}.
By (\ref{eqn:q-i-v-times-u-v-factored-into-product-pairs}),
we have for each pair $u, v$ of distinct elements of $V(G)$,
\[
p k_{u v} u_V = \sum_{i, j \in V(H)} \beta_{i j} q_i^u q_j^v u_V = \sum_{i, j \in V(H)} \beta_{i, j} \sum_{\substack{\phi \colon \phi(u) = i \\
\quad \phi(v) = j}} q_\phi = \sum_{\phi \in V(H)^V} \beta_{\phi(u), \phi(v)} q_\phi.
\]
Here to define the set of mappings $\phi$ we take $T = S \sqcup V$,
thus $V = T \setminus S$.
Then the last equality follows from the fact that
\[
\{ \phi : V \rightarrow V(H) \} = \bigsqcup_{i,j \in V(H)} \{ \phi \in V(H)^V : \phi(u) = i, \phi(v) = j \}
\]
is a partition.
Let $g = G + \calK$ be the corresponding element of $G$ in $\hat \calG(V)$.
Clearly $G = (\prod_{u v \in E(G)} K_{u, v}) U_V$ so $g = (\prod_{u v \in E(G)} k_{u v}) u_V$.
(When $E(G) = \emptyset$, we view it as $G = U_V$ so $g = u_V$.)
We have $f(G) = f(g)$ by the definition of $f$ on $\hat \calG$.
Also note that $g u_V = g$ so $g = g u_V^{|E(G)|}$.
Also, $p$ is an idempotent so $p = p^{|E(G)| + 1}$.
Then
\[
p g = p^{|E(G)| + 1} g u_V^{|E(G)|} = p \big( \prod_{u v \in E(G)} p k_{u v} u_V \big) u_V = p \big( \prod_{u v \in E(G)} ( \sum_{\phi \in V(H)^V} \beta_{\phi(u), \phi(v)} q_\phi ) \big) u_V.
\]
(When $E(G) = \emptyset$, this is simply $p g = p u_V$.)
Note that when we expand the product of sum
as a sum of products, for any two edges $uv \in E(G)$ and $u'v' \in E(G)$,
if the mappings $\phi$ and $\phi' \in V(H)^V$ (in the respective sums)
disagree on any vertex of $V = V(G)$, the product $q_\phi q_{\phi'} =0$.
This implies that in the sum of products expression
we only sum over all $\phi \in V(H)^V$ (and not over the $|E(G)|$-tuples of these).
Also $q_\phi$ resolves $p$, so $p q_\phi = q_\phi$.
Moreover, each $q_\phi \in \hat \calG(S \cup V)$ so $q_\phi u_V = q_\phi$.
This implies that
\[
p g = \big( \sum_{\phi \colon V \to V(H)} ( \prod_{u v \in E(G)} \beta_{\phi(u), \phi(v)} ) q_\phi \big) u_V = \sum_{\phi \colon V \to V(H)} ( \prod_{u v \in E(G)} \beta_{\phi(u), \phi(v)} ) q_\phi.
\]
(When $E(G) = \emptyset$, we view it as $p g = p u_V = \sum_{\phi \colon V \to V(H)} q_\phi = \sum_{\phi \colon V \to V(H)} ( \prod_{u v \in E(G)} \beta_{\phi(u), \phi(v)} ) q_\phi$ 
which is true by Claims~\ref{clm:idempotent-is-sum-of-its-resolving-basic-idempotents} and~\ref{clm:calP_(T,p)-resolved-by-q_phi} with $T = S \cup V$
and the fact that an element $h \in \hat \calG(S \cup V)$ resolves $p$ iff $h$ resolves $p u_V$.)
Note that $p \in \hat \calG(S)$
and has a representative in $\calG(S)$ as a 
linear combination of labeled graphs from $\calPLG(S)$,
$g \in \hat \calG(V)$ has the representative $G \in \calPLG(V) \subseteq \calG(V)$,
 and $S \cap V = \emptyset$.
Hence $f(p) f(g) = f(p g)$,  as $f$ is multiplicative, $f$ ignores labels and 
also by the definition of $f$ on $\hat \calG$.
Therefore by 
(\ref{f(q-phi)-factored-into-product-and-nonzero-1stpass})
and (\ref{eqn:q-phi-factored-into-product}),
\[
\begin{split}
f(p) f(G) &= f(p) f(g) = f(p g) = \sum_{\phi \colon V \to V(H)} ( \prod_{u v \in E(G)} \beta_{\phi(u), \phi(v)} ) f(q_\phi) \\
&= \sum_{\phi \colon V \to V(H)} ( \prod_{u v \in E(G)} \beta_{\phi(u), \phi(v)} ) ( \prod_{v \in V(G)} \alpha_{\phi(v)}) f(p),
\end{split}
\]
Since $f(p) \ne 0$, we can cancel it on both sides,
and complete the proof.

\begin{remark}
Note that if $D = 0$, then from the proof we get that $H$ is the empty graph,
so $f(G) = 0$ unless $G = K_0$ (the empty graph) and $f(K_0) = 1$.
After that, we trivially get $\hat \calG(T) = 0$ for any $T \ne \emptyset$
and $\hat \calG(\emptyset) \cong \bbF$ as algebras.
However, $p \in \calP_S$ so it follows that $S = \emptyset$.
Therefore by the previous isomorphism $\calP_\emptyset = \{ p \}$
so $p$ is the only basic idempotent in $\hat \calG(\emptyset)$ and so in the entire $\hat \calG$.
\end{remark} 
\end{proof}

%% file: extensions.tex
\section{Extensions}

So far we have allowed $G$ to have multiple edges but no loops
as is the standard definition. 
We can extend the results in this paper to more general graphs.
If we allow (multiple) loops in $G$,
we can show that 
the (multiplicative) graph parameter $f(G) = a^{\operatorname{\#loops}(G)}$
where $1 \ne a \in \bbF$ ($a$ can be $0$) cannot be expressed as a GH function,
even though its connection tensors $T(f, k, n)$ all have symmetric rank $1$ and $f(K_0) = 1$.
To get the corresponding representation theorem for graphs with (multiple) loops,
in the target graph $H$ each loop $e$ attached to a vertex $i$
must have two weights:
$\beta_{i i}$ which is used when a nonloop edge of $G$ is 
mapped onto $e$,  
and the other, say $\gamma_i$,
 when a loop of $G$ is mapped onto $e$.
In this extended model we have the following:
\begin{itemize}
\item The main expressibility results 
Theorems~\ref{thm:uniform-exponential-rank-boundedness}, \ref{thm:uniform-exponential-rank-boundedness-converse} and~\ref{thm:uniform-exponential-rank-boundedness-converse-strengthened}
remain true with the proof from Section~\ref{sec:proof-of-main-results}
carrying over to this model with slight adjustments.

\item The GH inexpressibility results from
Section~\ref{sec:apps}
remain true as the provided proofs involve only simple loopless graphs.
(For \#\textsc{Vertex-Disjoint-Cycle-Cover} ($\vdccover$),
a loop at a vertex is considered a cycle cover of that vertex;
this is consistent with the definition in Holant problems.)
%
%
%

\item The results from subsection~\ref{sec:bonded-deg-graphs} 
on bounded degree graphs remain true in the sense that
the inexpressibility results hold if we allow $\gamma_i$ to be arbitrary
(again, since only simple loopless graphs were used in the proof),
while the expressibility holds even with the stronger requirement $\gamma_i = \beta_{i i}$.
\end{itemize}


Analogously, a GH expressibility criterion can be stated
and proved within the framework of directed GH with minor adjustments, too.
We note that  generalizations of 
results in~\cite{Freedman-Lovasz-Schrijver-2007}
were given in~\cite{Lovasz-Schrijver-2008}
to a more general model which captures directed graphs, hypergraphs, etc.
We expect that it is possible to generalize
the GH expressibility criterion in this paper
for arbitrary fields
to this more general model
in a similar way as done in~\cite{Lovasz-Schrijver-2008}.

%% file: preliminary.tex
\section{Appendix}\label{sec:Preliminary-continued}


\subsection{Multilinear algebra}

We prove some statements we need about tensors. We assume that the reader is familiar with the definition of a multilinear function, 
tensor product, and dual space. It is good to start
with coordinate-free definitions because it allows a succinct notation.
But we will mostly use coordinates. 
The results are concrete, and they can be understood without too much
formalism.

Unless stated otherwise,
we do not impose a particular order
on the rows and columns
  of  matrices, or coordinates of  tensors.
The vector spaces may be infinite dimensional; and this infinite dimensionality is a
main technical point that causes some complications.

The tensor product of vector spaces $V_1, \ldots, V_n$ over $\bbF$ is denoted by $V_1 \otimes \cdots \otimes V_n$ or $\bigotimes_{i = 1}^n V_i$.
Elements of $\bigotimes_{i = 1}^n V_i$ are called order-$n$ tensors.
When $V_i = V$ for $1 \le i \le n$, 
we denote the tensor product by $V^{\otimes n}$.
(By convention $V^{\otimes 0} = \bbF$, and $v^{\otimes 0} = 1 \in \bbF$.)
Define a group action by
 $S_n$ 
on $V^{\otimes n}$ induced by
$\sigma(\otimes_{i = 1}^n v_i) = \otimes_{i = 1}^n v_{\sigma(i)}$.
Recall that $V^{\otimes n}$ consists of finite linear combinations of
such terms. 
We call a tensor $A \in V^{\otimes n}$ symmetric 
if $\sigma(A) = A$ for all $\sigma \in S_n$,
and denote by $\Sym^n(V)$ the set of symmetric tensors in $V^{\otimes n}$.
As $\bbF$ may have finite characteristic $p$,
the usual symmetrizing operator from $V^{\otimes n}$ to $\Sym^n(V)$,
which requires division by $n!$, 
is in general not defined.

Multilinear functions on $\prod_{i = 1}^n V_i$ can be naturally identified 
with the dual space $(\bigotimes_{i = 1}^n V_i)^*$ of 
linear functions on $\bigotimes_{i = 1}^n V_i$,
induced by  $f \mapsto f'$, satisfying $f'(\otimes_{i = 1}^n v_i) 
= f(v_1, \ldots, v_n)$.
Moreover, $\bigotimes_{i = 1}^n V_i^*$ canonically embeds into 
$(\bigotimes_{i = 1}^n V_i)^*$ via  
$(\otimes_{i = 1}^n f_i) (\otimes_{i = 1}^n v_i)
=
\prod_{i = 1}^n f_i(v_i)$. 
A special case is that $(V^*)^{\otimes n}$ embeds into $(V^{\otimes n})^*$.
If all $V_i$'s are finite 
dimensional then this embedding is an isomorphism. 
However, if $V_i$
are infinite dimensional, this embedding is \emph{not} surjective.
This was illustrated in subsection~\ref{Multilinear-algebra-short}.

Let  $A_i \colon V_i \to U_i$ 
($1 \le i \le n$) be linear maps  of vector spaces.
 They induce a homomorphism  $(\bigotimes_{i = 1}^n U_i)^* \to 
(\bigotimes_{i = 1}^n V_i)^*$ via $f \mapsto g$,
satisfying $g(\otimes_{i = 1}^n v_i) =
 f( \otimes_{i = 1}^n A_i v_i)$.

If $V_i$ are vector subspaces of $U_i$,  then $\bigotimes_{i=1}^n V_i$
canonically embeds in $\bigotimes_{i=1}^n U_i$.
In particular, if $V \subseteq U$, then $V^{\otimes n}$ and $\Sym^n(V)$
canonically embed in $U^{\otimes n}$ and $\Sym^n(U)$ respectively.
Under this embedding $\Sym^n(V) = \Sym^n(U) \cap V^{\otimes n}$.
We will also denote the space of symmetric 
$n$-fold multilinear functions on $V$ by $\Sym((V^{\otimes n})^*)$, 
i.e., the functions from $(V^{\otimes n})^*$ that are symmetric.
We have $(V^*)^{\otimes n} \cap \Sym((V^{\otimes n})^*) = \Sym^n (V^*)$.

In this paper, we are interested in vector spaces of the form
 $V = \bigoplus_{i \in {\calI}} \bbF_i$, 
or just $\bigoplus_{{\calI}} \bbF$,
where $\calI$ is an (index) set and each $\bbF_i,\, i \in {\calI}$, is a copy of $\bbF$ indexed by $i$.
In this case $V$ has a basis $\{e_i \mid  i \in {\calI}\}$,
and a vector $v \in V$ has finitely many nonzeros in this basis.
Note that for  infinite $\calI$ this is a proper subset of $\bbF^\calI$,
and in particular $\{e_i \mid  i \in {\calI}\}$
is \emph{not} a basis~\footnote{Of course every vector space 
has a basis; however this requires Zorn's Lemma so 
the proof is nonconstructive. In this paper our results are
constructive usually working with an explicitly given basis.} for  $\bbF^\calI$.
For $V = \bigoplus_{{\calI}} \bbF$, the dual
space $V^*$ can be 
identified with $\bbF^\calI$ via $f \mapsto (f(e_i))_{i \in {\calI}}$.
For  $V = \bigoplus_{{\calI}} \bbF$,
we have $V^{\otimes n} = \bigoplus_{{\calI}^n} \bbF$,
and $(V^{\otimes n})^*$ can be identified with
$\bbF^{\calI^n}$, the $n$-dimensional arrays.
We can view $\Sym(\bbF^{\calI^n}) = \Sym((V^{\otimes n})^*)$ as 
symmetric arrays, i.e., arrays in $\bbF^{\calI^n}$ that are 
invariant under permutations from $S_n$,
 with respect to the 
basis $\{e_i \mid  i \in {\calI}\}$ of $V = \bigoplus_{i \in \calI} \bbF_i$.


 
Any $A \in \bigotimes_{j = 1}^n V_j$, where $n \ge 1$, can be expressed as
a finite sum
\[
A = \sum_{i = 1}^r \bfv_{i 1} \otimes \cdots \otimes \bfv_{i n},
\quad \bfv_{i j}\in V_j.
\]
The least $r \ge 0$ for which $A$ has such an expression
is called the
rank of $A$,  denoted by $\rank(A)$.
$A=0$ iff $\rank(A)=0$.
If $r = \rank(A) > 0$ then in any such expression of $A$ of $r$ terms
all vectors
 $\bfv_{i j} \ne 0$.
When $n = 0$, $\bigotimes_{j = 1}^n V_j$ is $\bbF$,
and we define $\rank(A) = 1$ for $A \ne 0$ and $\rank(A) = 0$ for $A = 0$.

Similarly, for $A \in \Sym^n(V)$ we define the symmetric rank of $A$
to be the  least $r \ge 0$ for which $A$ can be expressed as
\[
A = \sum_{i = 1}^r \lambda_i \bfv_i^{\otimes n},
\quad \lambda_i \in \bbF, \bfv_i \in V,
\]
and is denoted by $\symrank(A)$.
If there is no such decomposition we define $\symrank(A) = \infty$. 
If $\symrank(A) < \infty$ then in any such expression of $A$
as a sum of $\symrank(A)$ terms
all $\lambda_i \ne 0$, all $\bfv_i \ne 0$ and are
pairwise linearly independent.
We show in Lemma~\ref{lm:finite-symmetric-tensor-rank} that
  for infinite $\bbF$,
$\symrank(A) < \infty$ for all $A \in \Sym^n(V)$.

We also need to refer to the rank of functions
in $(\bigotimes_{i = 1}^n V_i)^*$. As mentioned before 
$\bigotimes_{i = 1}^n V_i^*$ is embedded as a subspace 
of $(\bigotimes_{i = 1}^n V_i)^*$.
For a function $F \in (\bigotimes_{i = 1}^n V_i)^*$, where $n \ge 1$,
we define the rank of the function $F$ to be $\infty$ if $F \notin \bigotimes_{i = 1}^n V_i^*$,
and if $F \in \bigotimes_{i = 1}^n V_i^*$,
the rank of $F$ is the least $r$ for which $F$ can be written as
\[
F = \sum_{i = 1}^r \bff_{i 1} \otimes \cdots \otimes \bff_{i n},
\quad \quad \bff_{i j} \in V_j^*.
\]
When $n = 0$, $(\bigotimes_{j = 1}^n V_j)^*$ is $(\bbF)^* \cong \bbF$,
and we define $\rank(F) = 1$ for $F \ne 0$ and $\rank(F) = 0$ for $F = 0$.
The symmetric rank  $\symrank(F)$ of $F \in \Sym((V^{\otimes n})^*)$
  is similarly defined. 
It is $\infty$ if $F \not \in \Sym^n(V^*)$.
For $F \in \Sym^n(V^*)$, we define  $\symrank(F)$  to be the least $r$ 
such that
\[
F = \sum_{i = 1}^r \lambda_i \bff_i^{\otimes n},
\quad \quad \lambda_i \in \bbF, \quad \bff_i \in V^*,
\]
if such an expression exists; $\symrank(F) = \infty$ otherwise.
By the same Lemma~\ref{lm:finite-symmetric-tensor-rank}
for infinite $\bbF$, we have
$\symrank(F) < \infty$ for all $F \in \Sym^n(V^*)$.

Basically, the rank of a  multilinear  function is just an
extension of the tensor
rank from $\bigotimes_{i = 1}^n V_i^*$
 to $(\bigotimes_{i = 1}^n V_i)^*$.
Similarly the symmetric rank is the  extension 
from $\Sym^n(V^*)$ to $\Sym((V^{\otimes n})^*)$.
Clearly for all symmetric $A$,
$\rank(A) \le \symrank(A)$.
Both $\rank$ and  $\symrank$ are unchanged when moving from 
 $\bigotimes V_i$ to $\bigotimes U_i$,
if $V_i \subseteq U_i$. 





\begin{lemma}\label{lem:linear-independence-and-nonzero-minor-long}
The vectors $\bfx_1, \ldots, \bfx_r \in \bbF^\calI$ are linearly independent iff in the $r \times \calI$ 
matrix formed by $\bfx_1, \ldots, \bfx_r$ as rows 
there exists a nonzero $r \times r$ minor.
\end{lemma}
\begin{proof}
$\Leftarrow$ is obvious, so let us prove $\Rightarrow$.
Let $R \subseteq [r]$ be a maximal subset satisfying the property that
for some finite subset $C \subseteq \calI$ the
set of vectors $\{\bfx_i\mid_{C} \: : \, i \in R\}$ is linearly independent,
where $\bfx_i\mid_{C}$ is the restriction of $\bfx_i$ to $C$.
Suppose linear independence is achieved by $C$ for $R$.
Then
it also holds for any $C' \supseteq C$. 

If  $R \not = [r]$,  let $j \in [r] \setminus R$,
and consider $R^+ = R \cup \{j\}$. 
$\{\bfx_i\mid_{C}  \: : \, i \in R^+\}$ is linearly dependent.
Hence a unique linear combination holds for some $c_i \in \bbF$ ($i \in R$),
\begin{equation}\label{eqn:onC-lin-dep}
\bfx_j\mid_{C} \, = \sum_{i\in R} c_i \bfx_i\mid_{C}.
\end{equation}
For any $k \not \in C$,
$\{\bfx_i\mid_{C\cup\{k\}}  \: : \, i \in R^+\}$ is
also linearly dependent, and 
we have $\bfx_j\mid_{C\cup\{k\}}
= \sum_{i\in R} c'_i \bfx_i\mid_{C\cup\{k\}}$ for some $c'_i \in \bbF$.
Compared to (\ref{eqn:onC-lin-dep}), $c'_i = c_i$ for all $i \in R$.
Hence $\bfx_j \, = \sum_{i\in R} c_i \bfx_i$, a contradiction
to $\{\bfx_1, \ldots, \bfx_r\}$ being linearly independent.
So $R = [r]$.  There exists a nonzero $r \times r$ minor
in the $R \times C$ submatrix.
\end{proof}

For $\bfx = (x_i)_{i \in \calI} \in \bbF^\calI$
and  $h = (h_i)_{i \in \calI}  \in \bigoplus_\calI \bbF$ 
(in a direct sum, only finitely many $h_i$ are zero),
we denote their dot product
 by $\bfx(h) = \sum_{i \in \calI} x_i h_i \in \bbF$. 
 Here we view $\bbF^\calI$ as the dual space of $\bigoplus_\calI \bbF$. 
(In general the dot product of the pair  $\bfx, \bfy  \in \bbF^\calI$
is not defined.)

\begin{lemma}\label{lem:linear-independence-and-dual-vectors-long}
Let $\bfx_1, \ldots, \bfx_r \in \bbF^\calI$ be linearly independent. Then there exist $h_1, \ldots, h_r \in \bigoplus_{\calI} \bbF$ dual
to $\bfx_1, \ldots, \bfx_r$, i.e., $\bfx_i(h_j) = \delta_{i j},\, 1 \le i, j 
\le  r$.
\end{lemma}
\begin{proof}
By Lemma~\ref{lem:linear-independence-and-nonzero-minor-long}, there exist $r$ 
distinct indices $k_j \in \calI,\, 1 \le j \le r$ such that the matrix $A = (a_{i j})_{i, j = 1}^r = ((\bfx_i)_{k_j})_{i, j = 1}^r$ is invertible,
and let $B = (b_{i j}) = A^{-1}$.
Taking $h_i = \sum_{j = 1}^r b_{j i} e_{k_j} \in \bigoplus_\calI \bbF,\, 1 \le i \le r$, we see that the 
equality 
$AB = I_r$ directly translates into the desired result.
\end{proof}

\begin{lemma}\label{lem:quotient-space-dimension-long}
Let $\bfx_1, \ldots, \bfx_r \in \bbF^\calI$. Consider the linear map $\Phi \colon \bigoplus_\calI \bbF \to \bbF^r,\, h \mapsto ( \bfx_1(h), \ldots, \bfx_r(h) )$. Then $\dim (\bigoplus_\calI \bbF / \ker \Phi) = \dim \Span \{ \bfx_i \}_{i = 1}^r$.
\end{lemma}
\begin{proof}
By the First Isomorphism Theorem for vector spaces $\bigoplus_I \bbF / \ker \Phi \cong \im \Phi$.
So it suffices to prove $\dim \im \Phi = \dim \Span \{ \bfx_i \}_{i = 1}^r$. 
Clearly it suffices to prove the case when
$\bfx_1, \ldots, \bfx_r$ are linearly independent, and that follows
directly from Lemma~\ref{lem:linear-independence-and-dual-vectors-long}.
\end{proof}

%

\begin{lemma}\label{lem:tensor-powers-linear-independence-symmetric-rank-and-uniqueness-long}
Let $r \ge 0$, and
let $\bfx_1, \ldots, \bfx_r \in \bbF^\calI$ be $r$ linearly independent vectors and $a_1, \ldots, a_r \in \bbF \setminus \{0\}$. Then for any integer $n \ge 2$, the symmetric tensor
\begin{equation}\label{eqn:tensor-powers-linear-independence-symmetric-rank-and-uniqueness-decomposition-long}
A = \sum_{i = 1}^r a_i \bfx_i^{\otimes n} \in \Sym^n(\bbF^\calI)
\end{equation}
has $\symrank(A) = r$. For $n \ge 3$, any expression of $A$ as $\sum_{i = 1}^r b_i \bfy_i^{\otimes n}$ is a permutation of the sum in (\ref{eqn:tensor-powers-linear-independence-symmetric-rank-and-uniqueness-decomposition-long}).
\end{lemma}
\begin{proof}
When $r = 0$, the statement is trivially true so we assume $r \ge 1$.
Let $n \ge 2$ and $\symrank(A) = s$. Clearly $s \le r$. By being
of symmetric rank $s$, there
exist $\bfy_1, \ldots, \bfy_s \in \bbF^\calI$
 and $b_1, \ldots, b_s \in \bbF \setminus \{0\}$ such that
\begin{equation}\label{eqn:tensor-powers-linear-independence-symmetric-rank-and-uniqueness-lemma-two-sided-decomposition-long}
\sum_{i = 1}^r a_i \bfx_i^{\otimes n} = A = \sum_{j = 1}^s b_j \bfy_j^{\otimes n}.
\end{equation}
By Lemma~\ref{lem:linear-independence-and-dual-vectors-long},
there exist $h_1, \ldots, h_r$ dual to
$\bfx_1, \ldots, \bfx_r$.
For any $1 \le i \le r$,
applying $h_i^{\otimes (n-1)}$ to the sum, 
we get $a_i \bfx_i$ as a linear combination
of $\bfy_1, \ldots, \bfy_s$. 
Hence $s \ge r$ as $\bfx_1, \ldots, \bfx_r$ are linearly independent.
So $s = r$,
and $\bfy_1, \ldots, \bfy_s$ are linearly independent.

Next, let $n \ge 3$ and consider 
(\ref{eqn:tensor-powers-linear-independence-symmetric-rank-and-uniqueness-lemma-two-sided-decomposition-long}) again, where $s = r$.
Since $\symrank(A) = r$, all $b_j \ne 0$.
Applying $h_i$, we get
\begin{equation}\label{eqn:tensor-powers-linear-independence-symmetric-rank-and-uniqueness-lemma-expression-of-ai-xi-to-(n-1)-alt-long}
a_i \bfx_i^{\otimes (n - 1)}  = B = \sum_{j = 1}^r  b_j \bfy_j(h_i) \bfy_j^{\otimes (n - 1)}.
\end{equation}
From the LHS, $\symrank(B) = 1$.
 By what has just been proved,
 $\symrank(B)$ is the number of 
terms with nonzero coefficients on the RHS. 
Hence for any $i$, there is  exactly one $j$ such that
 $\bfy_j(h_i) \not = 0$.
Applying  $h_i^{\otimes (n-2)}$ to 
(\ref{eqn:tensor-powers-linear-independence-symmetric-rank-and-uniqueness-lemma-expression-of-ai-xi-to-(n-1)-alt-long}), we get
$a_i \bfx_i =  b'_j \bfy_j$,
where $ b'_j = b_j (\bfy_j(h_i))^{n-1} \not = 0$.
Since $\bfx_1, \ldots, \bfx_r$ are linearly independent, the map
$i \mapsto j$ is a permutation.
%
%
From $a_i \bfx_i =  b'_j \bfy_j$
we get $a_i = b'_j \bfy_j(h_i) = b_j (\bfy_j(h_i))^n$.
It follows that $\bfy_j = (a_i/b'_j)  \bfx_i = \bfy_j(h_i) \bfx_i$.
Therefore $b_j \bfy_j^{\otimes n} = b_j (\bfy_j(h_i))^n \bfx_i^{\otimes n} = a_i \bfx_i^{\otimes n}$.
Thus the expressions on LHS and RHS of~(\ref{eqn:tensor-powers-linear-independence-symmetric-rank-and-uniqueness-lemma-two-sided-decomposition-long})
are the same up to a permutation of the terms.
\end{proof}

\input{one-lemma-preliminary}

\input{finite-symmetric-tensor-rank}

\subsection{Subalgebras of \texorpdfstring{$\bbF^m$}{F\textasciicircum{m}}}\label{subsec:bbF^m-subalgebras}

%
%
%
We give a proof of Lemma~\ref{lem:F^m-subalgebras},
restated below.

\begin{lemma}\label{lem:F^m-subalgebras-long}
All subalgebras of $\bbF^m$, where $m \ge 0$, are of the following form:
For some partition $[m] = \bigsqcup_{i = 1}^s \calI_i$,
where $s \ge 0$, and $\calI_i \ne \emptyset$ for $i \in [s]$, 
the subalgebra has equal values on each $I_i$,
\[
\bbF^{(\calI_1, \ldots, \calI_s)} = \{ (c_1, \ldots, c_m) \in \bbF^m \mid \forall i \in [s], \ \forall j, j' \in \calI_i, \ c_j = c_{j'} \}.
\]
\end{lemma}
\begin{proof}
When $m = 0$, the statements is obvious.
Let $m \ge 1$ and $S \subseteq \bbF^m$ be a subalgebra of $\bbF^m$.
In particular, the multiplicative identity
is the $m$-tuple $(1, \dots, 1) \in S$.
We call $i, j \in [m]$ equivalent if $x_i = x_j$ for any $x = (x_1, \ldots, x_m) \in S$.
This is clearly an equivalence relation so
it partitions $[m]$ into (nonempty) equivalence classes $\calI_1, \ldots, \calI_s$
so that $[m] = \bigsqcup_{i = 1}^s \calI_i$.
As $[m] \ne \emptyset$ we have $s \ge 1$. 
We claim that $S = \bbF^{(\calI_1, \ldots, \calI_s)}$.
Clearly, $S \subseteq \bbF^{(\calI_1, \ldots, \calI_s)}$.
We prove the reverse inclusion.
For $s = 1$ this is clearly true
since the $m$-tuple $(1, \ldots, 1) \in S$
and $S$ is closed under scalar multiplication.

Now we let $s \ge 2$. By renaming  and omitting repeated coordinates
it is sufficient to prove the case when $m=s$ and  $\calI_i = \{i\}$.
Let $S' = \{(c_2, \ldots, c_s) \mid \exists c_1 \in \bbF,
(c_1, c_2, \ldots, c_s) \in S\}$ be the projection of $S$ to $\bbF^{s-1}$.
Clearly $S'$ is a subalgebra of $\bbF^{s-1}$, and 
by induction $S' = \bbF^{s-1}$.
Thus for some $b_1, \ldots, b_{s-1} \in \bbF$,
all  $s-1$ row vectors of the following $(s-1)\times s$
matrix $B$ belong to $S$, 
\begin{equation}\label{eqn:listing-elements-in-S'}
B = \left[
\begin{array}{ccccc}
b_1 & 1  &  0 & \ldots  & 0 \\
b_2 & 0  &  1 & \ldots  & 0 \\
\vdots & \vdots & \vdots & \ddots & \vdots \\
b_{s-1} &  0  &  0 & \ldots  & 1
\end{array}
\right].
\end{equation}
If all $b_i =0$, we let $v = (1, 1, \ldots, 1) \in S$.
Otherwise we may assume $b_1 \not =  0$.
By the definition of $I_1$ and $I_2$ and by the
closure of $S$ under scalar product and possibly adding $(1, 1, \ldots, 1)$,
for some $c_1 \not =  1$ and for some $c_3, \ldots, c_s \in \bbF$,
 we have $v' = (c_1, 1,  c_3, \ldots, c_s) \in S$.
Then multiplying $v'$ with the first row in (\ref{eqn:listing-elements-in-S'})
we get $v = (b_1c_1, 1, 0, \ldots, 0) \in S$.
Here $b_1c_1 \ne b_1$.
In either case, we obtain a matrix $A$ of rank $s$ by appending $v$  as
the last row to $B$.
Thus for all row
vectors $d \in \bbF^s$ the linear
system $x A = d$ has a solution $x \in \bbF^s$.
This shows that $d \in S$.
\end{proof}

%% file: one-lemma-preliminary.tex
\begin{lemma}\label{lem:tensor-powers-pairwise-implies-linear-independence-long}
Let $r \ge 0$, and
let $\bfx_1, \ldots, \bfx_r \in \bbF^\calI$ be $r$ nonzero
\emph{pairwise} linearly independent vectors. Then for any nonnegative integer $n \ge r-1$, the rank-$1$ symmetric tensors
\[
\bfx_1^{\otimes n}, \ldots, \bfx_r^{\otimes n} \in \Sym^n(\bbF^\calI)
\]
are linearly independent.
\end{lemma}
\begin{proof}
The case $r=0$ is vacuously true.
It is also trivially true for $r = 1$, 
since $\bfx_1^{\otimes n}$ is nonzero.
Assume $r \ge 2$. 
By pairwise linear independence, for every $1 \le i, j \le r$, $i \ne j$, 
from Lemma~\ref{lem:linear-independence-and-dual-vectors-long}
there exists $h_{i j}$ such that $\bfx_i(h_{i j}) = 1$, $\bfx_j(h_{i j}) = 0$. 
Suppose $\sum_{i = 1}^r \lambda_i \bfx_i^{\otimes n} = 0$ 
where $\lambda_i \in \bbF,\, 1 \le i \le r$. 
Applying $\bigotimes_{1 \le j \le r,\, j \ne i} h_{i j}$,
we get $\lambda_i \bfx_i^{\otimes (n - (r-1))} = 0$ for
$1 \le i \le r$, and thus $\lambda_i =0$ since $n \ge r-1$
and therefore $\bfx_i^{\otimes (n - (r-1))}$ is nonzero. 
Hence $\bfx_1^{\otimes n}, \ldots, \bfx_r^{\otimes n}$ are linearly independent.
\end{proof}

\begin{remark}
For $r \ge 2$,
the nonzero hypothesis is subsumed by
pairwise linear independence.
\end{remark}

%% file: finite-symmetric-tensor-rank.tex
\subsection{Finite symmetric tensor rank}\label{sec:fin-sym-ten-rank}
The proof of the following lemma is essentially the same as Lemma 4.2
in~\cite{Comon-Golub-Lim-Mourrain-2008};
the only modification needed is to avoid a
symmetrization step, which could result in a division by $0$
in a field of finite characteristic.

%
%
%
%
%
%
%
%
\begin{lemma}\label{lm:finite-symmetric-tensor-rank}
If $\bbF$ is a field of cardinality $|\bbF| > n$, a fortiori if 
$\bbF$ is infinite, and
$V$ is a vector space over  $\bbF$, then every symmetric
tensor $A \in  \Sym^n(V)$ has a  finite symmetric tensor rank
$\symrank(A) < \infty$.
Moreover, when $\dim V < \infty$,
we have $\symrank(A) \le \binom{\dim V + n - 1}{n}$.
\end{lemma}
\begin{proof}
By definition, every $A \in  \Sym^n(V) \subseteq V^{\otimes n}$ 
is a finite sum $A = \sum_{i=1}^m v_{i1} \otimes \cdots  \otimes
v_{in}$. Let $V' = {\rm span}\{v_{ij} \mid i \in [m], j \in [n]\}$
 be  a finite dimensional subspace of $V$.
As  $A \in  \Sym^n(V')$, we can assume
$V$ is finite dimensional, with no change in $\symrank(A)$; so we let $V = \bbF^N$, for some $N$.

Let $T =  {\rm span}\{ x^{\otimes n}  \mid x \in V\}
 \subseteq \Sym^n(V)$.  Our claim is that equality holds.
For every entry of $x^{\otimes n}$, which is  a product of
coordinate entries of $x$, we can classify it by how many factors are
the $j$-th coordinate of $x$, for $j \in [N]$. 
There are  $\binom{n + N - 1}{N - 1} = \binom{N + n - 1}{n}$ coordinates
 which can be indexed by tuples $(i_1, \ldots, i_N)$
where $i_1, \ldots, i_N \ge 0$ and $i_1 + \cdots + i_N = n$,
such that every entry of
every  $t \in T$ is equal to its entry at one of these coordinates.
We define a compression operator $C$ which selects only those
 $\binom{N + n - 1}{n}$ coordinates, and define
\[S =  {\rm span}\{ C(x^{\otimes n})  \mid x \in V\}
= \{ C(t)  \mid t \in T\}.\]

The compression operator $C$ is also applicable to $\Sym^n(V)$.
Indeed, by definition  as a symmetric array in $\bbF^{V^n}$,
 any $A \in \Sym^n(V)$ is invariant under any permutation of 
$n$. This means that
 for any $(k_1, \ldots, k_n)$, where $k_1, \ldots, k_n \in [N]$,
and any permutation $\pi \in S_n$,
if $(e^*_1, \ldots, e^*_{N})$ is the dual basis
to the canonical basis of $V$, then
$(e^*_{k_1} \otimes \cdots  \otimes e^*_{k_n}) (A)
= (e^*_{k_{\pi(1)}} \otimes \cdots  \otimes e^*_{k_{\pi(n)}}) (A)$.
This invariance can be characterized by the tuple
$(i_1,  \ldots,  i_N)$, where
$i_j = $ the number $j \in [N]$ among $(k_1, \ldots, k_n)$.
Thus  $C$ is applicable to $\Sym^n(V)$, and we denote
the result $C(\Sym^n(V)) = \{ C(v)  \mid v \in \Sym^n(V)\} \subseteq 
\bbF^{\binom{N + n - 1}{n}}$.
As $T \subseteq \Sym^n(V)$, we have $S \subseteq C(\Sym^n(V))$.

Next we prove that
$S = \bbF^{\binom{N + n - 1}{n}}$.
Then it follows that $S = C(\Sym^n(V))$, from which
it clearly follows that $T = \Sym^n(V)$ since $C$ simply
removes repeated entries.

Suppose otherwise, then $\dim S < \binom{N + n - 1}{n}$.
There exists a nonzero vector in $\bbF^{\binom{N + n - 1}{n}}$
such that it has a zero dot product with all $S$.
This means there exists a nonzero tuple
$(a_{(i_1,  \ldots,  i_N)}) \in \bbF^{\binom{N + n - 1}{n}}$
indexed by $N$-tuples of  nonnegative
integers that sum to $n$,
such that
$\sum_{(i_1,  \ldots,  i_N)} a_{(i_1,  \ldots,  i_N)} 
x_1^{i_1} \cdots x_N^{i_N} = 0$, for all $x_1, \ldots, x_N \in \bbF$.

As a polynomial in $x_N$ it has degree at most $n$, and yet it vanishes
at $|\bbF| > n$ points. So for any fixed $0 \le i_N \le n$,
$\sum_{(i_1,  \ldots,  i_{N-1})} a_{(i_1,  \ldots, i_{N-1},  i_N)}
x_1^{i_1} \cdots x_{N-1}^{i_{N-1}} = 0$, for all  $x_1, \ldots, x_{N-1}
 \in \bbF$, which can be viewed as a  polynomial in $x_{N-1}$ of
degree at most $n - i_N \le n$. 
Iterating $N$ steps, we reach a contradiction that
the tuple $(a_{(i_1,  \ldots,  i_N)})$ is entirely zero.

\end{proof}

%% file: appendix-previous-work.tex
\subsection{Some previous work and model distinction}\label{appendix-previous-work}
Schrijver~\cite{Schrijver-2009} gave a beautiful characterization
of a graph property $G \mapsto f(G)$ to be expressible as
$Z_H(\cdot)$ in (\ref{eqn:def-graph-homomorphismshomomorphisms-pa-fn})
with complex $\beta_{i,j}$ but all $\alpha_i=1$.
However, there are some subtle differences in the definition.
In particular, somewhat deviating from standard definition
(as in~\cite{Freedman-Lovasz-Schrijver-2007} and in this paper)
  the graphs $G$ are allowed to have loops (indeed multiloops
and multiedges) in~\cite{Schrijver-2009}. 
The criterion is expressed in terms of a sum $\sum_{P \in \Pi_{V(G)}}
\mu_P f(G/P) = 0$ for all $G$ with $|V(G)| > f(K_1)$. Here
 $\Pi_{V(G)}$  is the partition lattice on $V(G)$, $\mu_P$ is
the M\"{o}bius inversion function on partitions, and 
the graph $G/P$ is obtained from $G$ by condensing  all vertices 
in $P$ into one vertex. This condensation naturally creates loops.
In~\cite{Schrijver-2009} Schrijver carefully makes the distinction
between the results in that paper with that of~\cite{Freedman-Lovasz-Schrijver-2007},
and states that ``[A]n interesting question is how these results relate''.

In~\cite{Schrijver-2013} Schrijver gave another 
characterization of expressibility as
$Z_H(\cdot)$ in (\ref{eqn:def-graph-homomorphismshomomorphisms-pa-fn})
with complex $\beta_{i,j}$ but all $\alpha_i=1$.
Again, the graphs  $G$ may have multiloops
and multiedges. The criterion is expressed in terms of a  rank bound
of the $k$th ``connection matrix'' for every $k$.
However, here the definition of the ``connection matrix''
differs from that of~\cite{Freedman-Lovasz-Schrijver-2007}.
They are defined over ``$k$-marked  graphs'' where the $k$ marked vertices
are
not necessarily distinct. 
This is in contrast to
``$k$-labeled graphs'' in~\cite{Freedman-Lovasz-Schrijver-2007},
as well as in this paper.
Thus the  $k$th ``connection matrix'' in~\cite{Schrijver-2013}
is a super matrix of  the  $k$th ``connection matrix'' 
in~\cite{Freedman-Lovasz-Schrijver-2007}, and the  rank bound
is a stronger requirement.

At the end of this subsection we will show that this distinction
is material, by showing that the well-known \emph{hardcore gas} model
in statistical physics, $\sum_{\mbox{\scriptsize ind~} I \subseteq V(G)} \lambda^{|I|}$,
 where the sum is 
over all indepedent sets of $G$, cannot be expressed in the model
discussed in~\cite{Schrijver-2009,Schrijver-2013}, i.e., without
vertex weight. On the other hand, obviously the hardcore gas
model is defined as a partition function of GH \emph{with} vertex weight.
In particular the  rank bound for the
``connection matrix''  in~\cite{Schrijver-2013} must fail,
while it must hold for that of~\cite{Freedman-Lovasz-Schrijver-2007},
as well as for our connection tensor.

The terminology in the literature on this subject
is unfortunately not uniform. In Lov\'asz's book~\cite{Lovasz-book},
separate from the partition
function of GH as in (\ref{eqn:def-graph-homomorphismshomomorphisms-pa-fn}),
the model studied by B.~Szegedy in~\cite{B-Szegedy}
is called the ``edge coloring model''. This is
essentially what we called Holant problems, or edge models.
The slight difference is that  Holant problems allow
different 
 constraint functions from a  set  assigned at different vertices,
while the edge coloring model studied in~\cite{B-Szegedy}
is a special case of Holant problems where
for each arity $d$ a single symmetric vertex function  $f_d$
is given and  placed at all vertices of degree $d$.

In short, in edge models  edges play the role of
variables, and constraint functions are at vertices,
and in vertex models vertices play the role of
variables, and edges have binary constraint functions as well
as vertices have unary constraint functions.
Counting matchings, or perfect matchings, or valid
 edge colorings, or cycle covers
etc.\ are naturally expressible as edge models.
It turns out that many orientation problems can also be
expressed as edge models after a holographic transformation 
(by  $Z = \frac{1}{\sqrt{2}} \left[\begin{smallmatrix} 1 & 1\\
\mathfrak{i} & -\mathfrak{i} \end{smallmatrix}\right]$).
\emph{However}, confusingly,
the edge coloring model had also been called
a vertex model in~\cite{Draisma-et-al-2012}.
We also note that a preliminary version of~\cite{Schrijver-2015}
appeared as~\cite{Schrijver-2015-version1} which used
the terminology ``vertex model'', and that was changed to 
``edge coloring model'' in the final version~\cite{Schrijver-2015}. 
In particular, these papers~\cite{B-Szegedy,Draisma-et-al-2012,Schrijver-2015}
do not address the expressibility of counting perfect matchings
in the vertex model as 
in (\ref{eqn:def-graph-homomorphismshomomorphisms-pa-fn}),
which is the subject of the present paper.

Now we prove that the 
 hardcore gas
model fails the expressibility criterion of Schrijver~\cite{Schrijver-2013}
 as a partition function of GH \emph{without} vertex weight.
We observe that if we take $A =  \left[\begin{smallmatrix} 1 & 1\\
1 & 0 \end{smallmatrix}\right]$ indexed by $\{0, 1\}$,
which is the binary {\sc Nand} function representing the independent set
constraint, and the vertex weight $\alpha_0=1, \alpha_1 = \lambda$,
then  $\sum_{\mbox{\scriptsize ind~} I \subseteq V(G)} \lambda^{|I|}$
is clearly an instance of the expression
 (\ref{eqn:def-graph-homomorphismshomomorphisms-pa-fn}).
We now show that such an expression is impossible without vertex weight.

To state Schrijver's criterion we need a few definitions.
Let $f$ be a graph property. 
A $k$-marked graph is a pair $(G, \mu)$ where $G = (V, E)$ is graph,
and $\mu: [k] \rightarrow V$ is a function  marking $k$ (not necessarily
distinct)  vertices.
So a vertex may have several marks; 
this is the key distinction with $k$-labeled graphs.
For  $k$-marked graphs $(G_1, \mu)$ and $(G_2, \nu)$,
the product $(G_1, \mu) (G_2, \nu)$ is the $k$-marked graph 
by first taking a disjoint union and then merging all
 vertices of $G_1$ and $G_2$ that share a common mark.
(Note that this merging process may produce multiedges and multiloops.)
The $k$th \emph{connection matrix}  $C_{k, f}$  
 in the sense of~\cite{Schrijver-2013}
is the infinite matrix whose rows and columns are
indexed by all finite   $k$-marked graphs 
and the entry at $((G_1, \mu), (G_2, \nu))$
is $f((G_1, \mu) (G_2, \nu))$.
The expressibility criterion  by Schrijver in~\cite{Schrijver-2013}
 (Theorem 1) 
is  that $f(\emptyset) =1$ and for some $c$,
${\rm rank}(C_{k, f}) \le c^k$ for all $k$.

We show that for the  hardcore gas
model this rank grows super exponentially.
We consider the following submatrix. Let $\Pi_k$ be the
set of all partitions of $[k]$. The rows and columns of
the submatrix are indexed by the following  $k$-marked graphs.
For every $P = \{C_1, \ldots, C_s\} \in \Pi_k$
we have a  $k$-marked graph $G_P$ with $s$ vertices 
denoted by $v_1, \ldots, v_s$, with no edges. 
For every $j \in [k]$, let $C_t$ be the unique set
in $P$ that contains $j$, then we mark $v_t$ with $j$.
Now suppose $P, Q \in \Pi_k$ are the indices of a
row and column  respectively,
then the entry at $(P, Q)$ is the
 hardcore gas function evaluated at the product
graph $G_P G_Q$. We observe that this product
graph is just $G_{P \vee Q}$ where $P \vee Q$ is
 the least-upper-bound of $P$ and $Q$ in the lattice order of
refinement: $P \le P'$ iff $P$ refines $P'$.
It follows that the entry at  $(P, Q)$ is
\[\prod_{D \in P \vee Q } ( 1 + \lambda )
= ( 1 + \lambda )^{| P \vee Q |}.\]
It is proved in~\cite{Schrijver-2013} (Proposition 1)
for $\lambda  \not = -1, 0, 1, 2, ..., k-2$,
this matrix is nondegenerate. 
As its  dimension $|\Pi_k|$
grows superexponentially in $k$,
we see that the criterion in~\cite{Schrijver-2013} is not satisfied.